\documentclass[pdflatex,sn-mathphys-num]{sn-jnl}% Math and Physical Sciences Numbered Reference Style
\usepackage{graphicx}%
\usepackage{multirow}%
\usepackage{amsmath,amssymb,amsfonts}%
\usepackage{amsthm}%
\usepackage[mathscr]{eucal}%
\usepackage[title]{appendix}%
\usepackage{caption}
\usepackage{xcolor}%
\usepackage{textcomp}%
\usepackage{manyfoot}%
\usepackage{algorithm}%
\usepackage{algorithmicx}%
\usepackage{algpseudocode}%
\usepackage{listings}%
\usepackage{subcaption}
\usepackage{multirow}
\usepackage{bbm}
\usepackage{tabularx,booktabs,array,adjustbox}
\newcolumntype{L}[1]{>{\raggedright\arraybackslash}p{#1}}
\PassOptionsToPackage{compatibility=false}{caption}
\usepackage[svgnames]{xcolor}
\hypersetup{
	colorlinks=true,
	citecolor=Green,
	linkcolor=Red,
	urlcolor=blue}
\hypersetup{bookmarksdepth=3}
\graphicspath{{./Images/}}
\usepackage{tikz}
\usetikzlibrary{positioning,decorations.pathreplacing, automata, arrows.meta,calc}
\usepackage{pgfplots}
\pgfplotsset{compat=1.18}

\DeclareMathOperator{\EX}{\mathbb{E}}% expected value
\DeclareMathOperator*{\argmax}{argmax}% thin space, limits underneath in displays

\theoremstyle{thmstyleone}%
\theoremstyle{thmstyletwo}%

\theoremstyle{thmstylethree}%
\usepackage{rotfloat} 
\makeatletter
\def\fps@sidewaystable{tbp}
\def\fps@sidewaysfigure{tbp}
\makeatother

\begin{document}

\title[Article Title]{A Review On Safe Reinforcement Learning Using  Lyapunov and Barrier Functions}


\author*[1]{\fnm{Dhruv} \sur{Kushwaha}}\email{dhruv.kushwaha@ufl.edu}

\author[1]{\fnm{Zoleikha} \sur{Biron}}\email{z.biron@ece.ufl.edu}

\affil[1]{\orgdiv{Department of Electrical \& Computer Engineering}, \orgname{University of Florida}, \orgaddress{\street{968 Center Dr}, \city{Gainesville}, \postcode{32603}, \state{FL}, \country{USA}}}


\abstract{Reinforcement learning (RL) has proven to be particularly effective in solving complex decision-making problems for a wide range of applications. From a control theory perspective, RL can be considered as an adaptive optimal control scheme. Lyapunov and barrier functions are the most commonly used certificates to guarantee system stability for a proposed/derived controller and constraint satisfaction guarantees, respectively, in control-theoretic approaches. However, compared to theoretical guarantees available in control-theoretic methods, RL lacks closed-loop stability of a computed policy and constraint satisfaction guarantees. Safe reinforcement learning refers to a class of constrained problems where the constraint violations lead to partial or complete system failure. The goal of this review is to provide an overview of safe RL techniques using Lyapunov and barrier functions to guarantee this notion of safety (stability of the system in terms of a computed policy and constraint satisfaction during training and deployment). Three concrete takeaways emerge from our analysis: (i) the field has shifted decisively from model-based to model-free formulations since 2017, with combined CLF--CBF approaches becoming the most active sub-area post-2022; (ii) per-class open problems are now well-defined, certificate validity under function approximation and distribution shift for Lyapunov methods, feasibility and deadlock under hard CBF--QP shielding for barrier methods, and joint CLF--CBF feasibility under model uncertainty for combined methods; and (iii) deployment to high-dimensional and partially observable settings remains the dominant scalability barrier across all three classes. The different approaches employed are discussed in detail along with their shortcomings and benefits to provide critique and possible future research directions. The review demonstrates promising scope for providing safety guarantees for complex dynamical systems with operational constraints using model-based and model-free RL.}




\maketitle

\section{Introduction}\label{sec1}

With increases in efficiency and function approximation capabilities of deep neural networks (DNN), reinforcement learning has seen increased research and some exciting developments over the last few decades~\cite{arulkumaran2017deep}. RL in its bare bones involves an agent iteratively interacting with an environment to compute a control or decision making policy by maximizing a reward function. RL has proven to be effective in computing control or decision policies for a wide variety of complex systems and environments such as robotics \cite{ibarz2021train, wang2022distributed, abeyruwan2023sim2real, liu2022digital, muzio2022deep}, computer vision \cite{pinto2023tuning, le2022deep, tao2022evaluating}, cyber-security \cite{adawadkar2022cyber, piplai2022knowledge, tran2022cascaded, huang2022reinforcement, 10252253}, energy management \cite{ganesh2022review, liu2023energy, yang2022reinforcement, 9916919}, chess \cite{bertram2022supervised, xu2022data, hammersborg2022reinforcement} and video games \cite{taylor2014reinforcement, mnih2013playing, kaiser2019model} to name a few. However, when using RL there are several complications and theoretical gaps in terms of reproducibility, convergence guarantees, large amounts of data for training and large number of iterations required for convergence~\cite{henderson2018deep}. Even though RL is often known for its data-driven nature, it can be broadly classified into two approaches for deriving a control or decision policy namely, model-based (value and policy iteration) and model-free (value-function, policy search or gradient and actor-critic methods) approaches. Model-based approaches imply the use of knowledge (partial or complete) or an approximation of the system to derive a control policy whereas, model-free approaches purely rely on data or experiences collected by interactions with the environment without using a model of the system~\cite{osinenko2022reinforcement}. It seems intuitive to have stronger guarantees for constraint satisfaction and computing a stabilizing policy while leveraging a model-based approach as the RL agent uses information from the system model. There has been some success in providing these guarantees for model-free approaches as well, notably in~\cite{cheng2019end}. \em A key challenge for policies derived through use of RL has been providing guarantees for stability of the system for a computed policy and satisfying system state constraints. \em Broadly speaking this challenge encapsulates the need for safety in RL for real-world applications and thus, encourages the research in this field.

There are several methods in the literature for computing a stabilizing policy and satisfying system constraints in RL. The simplest approaches offer a direct and intuitive method of filtering out ``unsafe'' actions and policies. The term ``unsafe'' can be referred to any action leading to constraint violation or system instability. In~\cite{curi2022safe}, the authors propose a confidence safety filter for stochastic nonlinear systems. The key idea presented is to formulate safety constraints as cumulative costs which can then be expressed as cost constraints for RL. Backup policies in case of constraint violations are obtained for the safety filter by computing a safe policy through robust RL, which is then employed in a confidence-based safety filter. The authors in~\cite{vinod2022safe} use a convex optimization-based filtering to satisfy hard constraints for target tracking. They propose a two-level motion planner where a RL-based controller generates an input based on the target and quadratic programming (QP) based safety filter generates a safe input, regulation is performed on these two inputs produced. Similar to safety filter, supervisor-based safe RL methods often use human supervision~\cite{arakawa2018dqn, wu2023toward, luo2023human} to penalize or replace unsafe actions with desirable ones. These supervisor-based methods are often termed as human-in-loop RL. A formal approach to supervisor-based safe RL methods are classified as RL via shielding. Authors in~\cite{elsayed2021safe} propose a centralized (single shield for all agents in an environment) and factored (separate shields for a subsets of agents in an environment) shielding methodology for a multi-agent RL setting using linear temporal logic (LTL) for safety constraints. The scalability and learning performance are discussed for each methodology by the authors of mentioned work. In~\cite{alshiekh2018safe}, authors express constraints as temporal logic and propose an algorithm for automatic synthesis of shields for the given temporal logic. This is further incorporated with RL and the results show increased learning performance compared to unshielded cases. However, a key drawback of this approach is that it relies on some prior information for model of the system to deem an action unsafe. 
Model predictive control (MPC) has long been effective in control theory for handling uncertainty in dynamical systems while satisfying operational constraints. There have been several notable works in combining the data-driven advantage of RL and the robustness of MPC, reviewed in~\cite{norouzi2023integrating}. RL using robust MPC is proposed in~\cite{zanon2020safe}, where the authors use online data to compute safe design constraints to evaluate the cost function for RL, Q-function and value function are obtained via robust MPC. In~\cite{sawant2022bridging}, combination of QP and MPC, one at a time with RL is explored. The policy and value functions of RL are approximated using MPC and QP instead of employing DNNs. Compared to lack of guarantees and explainability of DNNs, this approach provides structure to the approximations for policy and value functions.

Control Lyapunov functions (CLF)~\cite{sontag1999control} and control barrier functions (CBF)~\cite{ames2019control} have proven to be an effective way to guarantee safety through stability of a closed-loop system and defining safe sets, respectively. Ensuring safe operation for dynamical systems involving human interaction, expensive equipment etc., requires strong guarantees during and after the learning process for RL agents. Using CBFs and CLFs to provide stability and safety guarantees is especially well suited for learning because both features may be stated using learnable functions. \em Considering their usefulness in a learning framework and their applicability to a wide variety of control frameworks, this paper reviews methods using Lyapunov and barrier functions to ensure safety and stability guarantees of a computed policy while using RL for decision-making problems. \em Furthermore, most real-world applications require safety during the training process (for example, navigation in unknown environments for drones, human presence in application etc.) which makes this problem crucial as it provides theoretical guarantees for safety. \em The key motivation for this review is to discuss methods employed in recent work, not covered in reviews and surveys mentioned in Section~\ref{section:Reviewsandsurveys} and particularly focus on use of Lyapunov and barrier functions for RL to guarantee satisfying system constraints and stability in terms of the computed policy. \em We focus more on the theoretical approaches present in the literature rather than the type of applications it is employed. 

Rest of the review is organized as follows: Section~\ref{section:Reviewsandsurveys} presents scope of work for different reviews and surveys considered. Section~\ref{section:TechnicalBackground} discusses preliminary theory and notations briefly to provide the reader a background for further discussion. Section~\ref{section:ControlLF}, \ref{section:ControlBF} and \ref{section:CLFandCBF} summarizes, classifies and discusses the current and prior work related to use of: (i) Lyapunov functions in RL; (ii) Barrier functions in RL; (iii) Lyapunov and barrier functions in RL, respectively. Section~\ref{section:Discussions} provides a discussion on simulation benchmarks, theoretical developments, convergence properties, soft vs hard safety constraints data efficiency, future research directions, current challenges and potential real-world applications. Finally, Section~\ref{section:Conclusions} presents conclusions to the review conducted. 
%%%%%%%%%%%%%%%%%%%%%%%%%%%%%%%%%%%%%%%%%%%%%%%%%%%%%%%%%%%%%%%%%%%%%%%%%%%%%%%%%%%%%%%%%%%%%%%%%%%%%%%%%%%%%%%%%%%
%%%%%%%%%%%%%%%%%%%%%%%%%%%%%%%%%%%%%%%%%%%%%%%%%%%%%%%%%%%%%%%%%%%%%%%%%%%%%%%%%%%%%%%%%%%%%%%%%%%%%%%%%%%%%%%%%%%
\section{Related Work}\label{section:Reviewsandsurveys}
The literature in safe reinforcement learning has been fairly recent, spanning the last two decades. The earliest survey in this field by~\cite{garcia2015comprehensive}, primarily focuses on safe RL by augmenting the optimization criterion and exploration process by including a risk parameter and state constraints. The survey also focuses on leveraging external knowledge in the form of expert policies and human intervention to guide the exploration process. The review by Liu et al.~\cite{liu2021policy} focuses on policy learning with constraints for model-free RL. The authors present a robust taxonomy for the types of constraints for a constrained Markov decision process (CMDP) and discusses various policy optimization approaches based on the type of constraints in CMDP. The authors broadly classify constraints as cumulative and instantaneous depending on the long-term and short-term effects of constraints on the cost function, respectively.

\begin{sidewaystable}[htpb]
\centering
\caption{Contribution of this review compared with prior surveys discussed in Section~\ref{section:Reviewsandsurveys}.
Coverage scale: $\bullet\bullet\bullet$ comprehensive (dedicated section with method classification);
$\bullet\bullet\circ$ substantial (sub-section / multi-paragraph);
$\bullet\circ\circ$ brief (few instances);
$\circ\circ\circ$ absent, out of declared scope, or pre-dates the literature.}
\label{tab:review_comparison}
\footnotesize
\setlength{\tabcolsep}{4pt}
\renewcommand{\arraystretch}{1.15}
\begin{tabular}{@{}p{2.8cm} c p{3.6cm} c c c c c c c c@{}}
\toprule
\textbf{Review} & \textbf{Yr} & \textbf{Declared Scope}
& \shortstack{\textbf{Lyap}\\\textbf{in RL}}
& \shortstack{\textbf{CBF}\\\textbf{in RL}}
& \shortstack{\textbf{CLF+CBF}\\\textbf{in RL}}
& \textbf{CMDP}
& \shortstack{\textbf{Bench-}\\\textbf{marks}}
& \shortstack{\textbf{2023--25}\\\textbf{Lit.}}
& \shortstack{\textbf{Open}\\\textbf{Prob.}}
& \shortstack{\textbf{Apps}\\~} \\
\midrule
Garc\'ia \& Fern\'andez~\cite{garcia2015comprehensive} & 2015
& Risk-augmented exploration; external knowledge
& $\circ\circ\circ$ & $\circ\circ\circ$ & $\circ\circ\circ$
& $\bullet\circ\circ$ & $\circ\circ\circ$ & $\circ\circ\circ$
& $\bullet\circ\circ$ & $\bullet\circ\circ$ \\

Liu et al.~\cite{liu2021policy} & 2021
& Constrained policy learning, model-free RL
& $\circ\circ\circ$ & $\circ\circ\circ$ & $\circ\circ\circ$
& $\bullet\bullet\bullet$ & $\bullet\circ\circ$ & $\circ\circ\circ$
& $\bullet\circ\circ$ & $\bullet\circ\circ$ \\

Anand et al.~\cite{anand2021safe} & 2021
& Safe learning (RL+SL) using CLF/CBF
& $\bullet\bullet\circ$ & $\bullet\bullet\circ$ & $\bullet\circ\circ$
& $\bullet\circ\circ$ & $\circ\circ\circ$ & $\circ\circ\circ$
& $\bullet\circ\circ$ & $\bullet\circ\circ$ \\

Brunke et al.~\cite{brunke2022safe} & 2022
& Safe learning for robotics (bird's-eye)
& $\bullet\bullet\circ$ & $\bullet\bullet\circ$ & $\bullet\circ\circ$
& $\bullet\bullet\circ$ & $\bullet\bullet\circ$ & $\circ\circ\circ$
& $\bullet\bullet\circ$ & $\bullet\bullet\bullet$ \\

Gu et al.~\cite{gu2022review} & 2022
& Safe RL theory and applications; 2H3W; MARL
& $\bullet\circ\circ$ & $\bullet\circ\circ$ & $\bullet\circ\circ$
& $\bullet\bullet\circ$ & $\bullet\bullet\circ$ & $\circ\circ\circ$
& $\bullet\bullet\circ$ & $\bullet\bullet\circ$ \\

Osinenko et al.~\cite{osinenko2022reinforcement} & 2022
& RL with guarantees (supervisor / MPC / Lyapunov)
& $\bullet\bullet\circ$ & $\bullet\circ\circ$ & $\bullet\circ\circ$
& $\bullet\circ\circ$ & $\circ\circ\circ$ & $\circ\circ\circ$
& $\bullet\circ\circ$ & $\bullet\circ\circ$ \\

Dawson et al.~\cite{dawson2022safe} & 2022
& Neural certificates for control (CLF/CBF/contraction)
& $\bullet\bullet\circ$ & $\bullet\bullet\circ$ & $\bullet\bullet\circ$
& $\bullet\circ\circ$ & $\bullet\circ\circ$ & $\circ\circ\circ$
& $\bullet\bullet\circ$ & $\bullet\circ\circ$ \\

\midrule
\textbf{This review} & \textbf{2026}
& \textbf{Lyapunov and barrier functions in RL specifically}
& $\boldsymbol{\bullet\bullet\bullet}$ & $\boldsymbol{\bullet\bullet\bullet}$ & $\boldsymbol{\bullet\bullet\bullet}$
& $\boldsymbol{\bullet\bullet\circ}$ & $\boldsymbol{\bullet\bullet\circ}$ & $\boldsymbol{\bullet\bullet\bullet}$
& $\boldsymbol{\bullet\bullet\bullet}$ & $\boldsymbol{\bullet\bullet\circ}$ \\
\bottomrule
\end{tabular}
\end{sidewaystable}

Brunke et al.~\cite{brunke2022safe} provides an extensive and one of the most comprehensive reviews covering safe learning for control from a robotics perspective, however, they do not cover key recent and prior works on combining Lyapunov and barrier functions in RL. To quote the authors, they provide a ``bird's eye view'' of the field. Authors cover most of the literature associated with safe learning and provide excellent categorizations for levels of safety and various formulations for safe learning techniques. However, due to the vast range of topics considered, use of Lyapunov and barrier functions in RL is sparsely covered. The review does provide an excellent starting point for an overview into all possible directions for development in safe learning for control and RL. Gu et al.~\cite{gu2022review} focus on all developments in safe RL covering theory and applications. The authors formalize a ``2H3W'' problem, addressing the key problems and challenges pertaining to safe RL implementation. Convergence analysis and iteration complexity for various approaches to safe RL are explored for model-free and model-based RL along with applications and available benchmarks. The authors also discuss application of safe RL in multi-agent settings, discussing possible research avenues. The review is detailed, however, it covers only a few approaches pertaining to use of Lyapunov and barrier functions.

The authors in~\cite{osinenko2022reinforcement} present an overview for RL with guarantees. The authors cover three approaches for stability of the system for a computed policy and constraint satisfaction namely, RL with supervisors, model predictive control (MPC)-based RL and Lyapunov-based RL. The review presents few instances of using CBF with CLF, but it primarily gives an overview of key challenges in consideration of model-based and model-free RL while providing safety guarantees. Anand et al.~\cite{anand2021safe} present a concise and focused review on safe learning for control using CLF and CBF. The authors cover safety in RL, online supervised learning (SL) and offline SL with incorporation of CLF, CBF and combining CLF with CBF. However, the review presents a few notable approaches for combining CLF and CBF with RL and only covers one CMDP based formulation. The excellent survey by Dawson et al.~\cite{dawson2022safe} covers a great depth of literature available on learning-based control using CLF, CBF and contraction methods. The review provides excellent theoretical overview of different formulations and implementations using control certificates but is more focused towards learning-based control (adaptive controls) from a control-theoretic perspective compared to RL.

Table~\ref{tab:review_comparison} positions this review against the seven prior surveys discussed above along nine axes. Three contribution gaps motivate the present work. First, prior reviews treat the combined use of CLF and CBF in RL only briefly (Anand et al.) or from a control-theoretic perspective with limited RL emphasis (Dawson et al.). We devote a dedicated Section \ref{section:CLFandCBF} with a method classification Table~\ref{tab:lyap_barrier_adv_disadv}. Second, every prior review cuts off at 2022 or earlier, leaving 2023--2025 contributions on data-driven CLBFs, neural Lyapunov–barrier verification, disturbance-observer CBFs, and NODE-based safe RL outside their scope. Third, no prior review provides per-method-class open-problem analysis grounded in the specific failure modes of each formulation (filter feasibility, drift estimation, certificate validity under approximation, joint CLF–CBF infeasibility), which we provide in Section \ref{subsec:lyap_discussion_open_problems}, \ref{subsec:cbf_discussion_open_problems}, and \ref{subsec:clbf_open}.

The goal of this review is to primarily focus on safe RL using Lyapunov and barrier functions to provide an overview on recent and prior important developments in the field along with possible future research directions. To the best of author's knowledge, a current review/survey on recent advances is not present in the literature, which is addressed in this review. To clarify the scope, we position this work as an RL-centric synthesis rather than a general survey of safe learning for control. In contrast to broad surveys on safe learning and control certificates, the organizing principle here is how Lyapunov and barrier functions are embedded into the RL loop: as rewards, critics, losses, constraints, safety filters, or verification objects.

\subsection{Review Methodology and Scope}\label{subsection:review_methodology}
The literature was collected through a narrative, semi-systematic search over IEEE Xplore, ACM Digital Library, ScienceDirect, SpringerLink, PMLR/OpenReview proceedings, arXiv, and Google Scholar. Search terms included combinations of ``safe reinforcement learning'', ``control Lyapunov function'', ``Lyapunov reinforcement learning'', ``control barrier function'', ``barrier function reinforcement learning'', ``CLF-CBF-QP'', ``safe actor-critic'', ``safe adaptive dynamic programming'', ``offline safe reinforcement learning'', and ``safe RL benchmark''. We included papers that explicitly integrate Lyapunov functions, control Lyapunov functions, barrier functions, control barrier functions, Lyapunov--barrier functions, or related certificate conditions into an RL/ADP learning loop. Papers were excluded when they used only generic constrained RL without certificate structure, pure model predictive control without RL/ADP learning, or purely supervised certificate learning without a connection to policy learning or policy filtering.

The reviewed works were categorized according to two criteria. First, the certificate type: Lyapunov-only, barrier-only, or combined Lyapunov--barrier. Second, the role of the certificate in the algorithm: reward/cost shaping, critic or value-function parameterization, actor/critic loss regularization, policy projection, QP-based safety filtering, formal verification, or bilevel/constrained optimization. Because safe RL is rapidly evolving, recent preprints and early-access articles are included when they directly address the certificate-based safe RL theme; these entries are treated as emerging directions rather than settled consensus.
%%%%%%%%%%%%%%%%%%%%%%%%%%%%%%%%%%%%%%%%%%%%%%%%%%%%%%%%%%%%%%%%%%%%%%%%%%%%%%%%%%%%%%%%%%%%%%%%%%%%%%%%%%%%%%%%%%%
%%%%%%%%%%%%%%%%%%%%%%%%%%%%%%%%%%%%%%%%%%%%%%%%%%%%%%%%%%%%%%%%%%%%%%%%%%%%%%%%%%%%%%%%%%%%%%%%%%%%%%%%%%%%%%%%%%%
\section{Theoretical Background}\label{section:TechnicalBackground}
Some formal definitions and notations are described in this section to give context for further discussion. The theory is kept brief and sources for detailed explanations are cited.

\subsection{Notation}\label{subsec:notation}
To support a unified treatment across control-theoretic and RL formulations, we use the symbol mapping summarized in Table~\ref{tab:notation}. Throughout the paper, $x \in \mathcal{X} \subseteq \mathbb{R}^n$ denotes the continuous-time state used in dynamical-system formulations (Sections~\ref{subsection:SafetyCriterion},~\ref{subsection:CLF},~\ref{subsection:CBF}) and $s \in \mathcal{S}$ denotes the discrete MDP state (Section~\ref{subsection:CMDP}); these refer to the same underlying physical state under a sampling discretization $\Delta t$. Similarly, $u \in \mathcal{U}$ and $a \in \mathcal{A}$ denote the continuous control input and discrete MDP action respectively.

\begin{table}[htpb]
\centering
\caption{Unified notation used throughout the paper.}\label{tab:notation}
\small
\begin{tabular}{@{}ll@{}}
\toprule
Symbol & Meaning \\
\midrule
$x \in \mathcal{X} \subseteq \mathbb{R}^n$ & Continuous-time state \\
$s \in \mathcal{S}$ & Discrete MDP state ($s = x$ at sampling instants) \\
$u \in \mathcal{U} \subseteq \mathbb{R}^{m_u}$ & Continuous control input \\
$a \in \mathcal{A}$ & MDP action ($a = u$ at sampling instants) \\
$f(x,u)$ & State-transition function (continuous time) \\
$f_{cl}(x)$ & Closed-loop dynamics under a fixed policy \\
$\mathbb{P}(s'|s,a)$ & MDP transition probability \\
$\gamma \in (0,1]$ & MDP discount factor \\
$\eta \geq 0$ & Barrier damping coefficient (formerly $\gamma$ in Eq.~(19b)) \\
$\mathcal{R}, \mathcal{C}$ & MDP reward and constraint-cost functions \\
$\pi$ & Policy; $\pi^*$ optimal policy \\
$V(x)$ & Lyapunov function / state-value function \\
$Q(s,a)$ & State-action-value function \\
$h(x)$ & Barrier function; $\mathcal{C} = \{x : h(x) \leq 0\}$ safe set \\
$\alpha(\cdot)$ & Extended class-$\mathcal{K}$ function \\
$L_f, L_g$ & Lie derivatives along $f$ and $g$ \\
$\mathcal{X}_s, \mathcal{X}_u, \mathcal{X}_f$ & Safe set, unsafe set, largest feasible set \\
$\delta \geq 0$ & Slack variable in relaxed CLF--CBF--QP \\
\bottomrule
\end{tabular}
\end{table}

\subsection{Definition of Safety}\label{subsection:SafetyCriterion}
A general continuous time nonlinear dynamical system is represented as $\dot{x} = f(x,u)$, where $x \in \mathcal{X} \subseteq \mathbb{R}^n$ is the state of the system, $u \in \mathcal{U} \subseteq \mathbb{R}^m$ is the control input and $f: \mathcal{X} \times \mathcal{U} \mapsto \mathcal{X}$ is the state transition function (assumption: Locally Lipschitz in $x$ and $u$) \cite{dawson2022safe}. The control affine form of nonlinear dynamical system can represented as $\dot{x} = f(x) + g(x)u$, where $g:\mathbb{R}^n \mapsto \mathbb{R}^{n \times m}$ (assumption: Locally Lipschitz in $x$ and $u$).

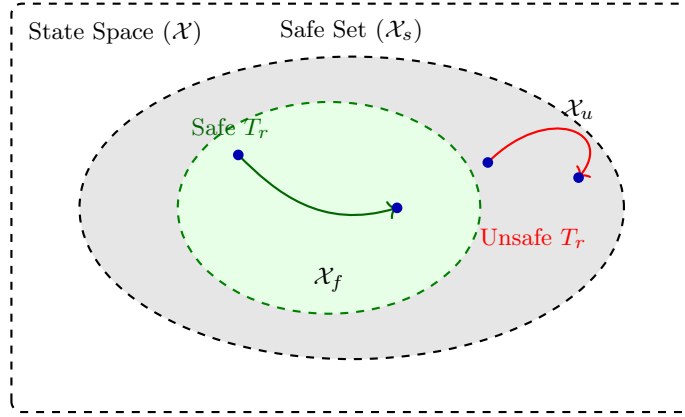
\begin{figure}
    \centering
\begin{tikzpicture}[scale=1.0, every node/.style={font=\small}]
  % outer state space
  \draw[dashed, thick, rounded corners=4pt] (-4.5,-2.7) rectangle (4.5,2.7);
  \node[anchor=north west] at (-4.4,2.6) {State Space ($\mathcal{X}$)};

  % unsafe set X_u (large dashed ellipse)
  \draw[dashed, thick, fill=gray!20] (0,0) ellipse (3.6 and 2);
  \node at (3.0,1.3) {$\mathcal{X}_u$};
  \node[anchor=south] at (0,2.05) {Safe Set ($\mathcal{X}_s$)};

  % feasible set X_f (inner dashed green ellipse)
  \draw[dashed, thick, green!50!black, fill=green!10] (-0.3,0) ellipse (2 and 1.4);
  \node at (-0.3,-0.95) {$\mathcal{X}_f$};

  % safe trajectory
  \draw[thick, green!40!black, ->] (-1.5,0.7) .. controls (-1.0,0.2) and (-0.4,-0.3) .. (0.6,0.0);
  \fill[blue!70!black] (-1.5,0.7) circle (2pt);
  \fill[blue!70!black] (0.6,0.0) circle (2pt);
  \node[green!40!black] at (-1.6,1.05) {Safe $T_r$};

  % unsafe trajectory crossing into X_u
  \draw[thick, red, ->] (1.8,0.6) .. controls (2.6,1.4) and (3.5,1.0) .. (3.0,0.4);
  \fill[blue!70!black] (1.8,0.6) circle (2pt);
  \fill[blue!70!black] (3.0,0.4) circle (2pt);
  \node[red] at (2.4,-0.4) {Unsafe $T_r$};
\end{tikzpicture}
\caption{Safe set ($\mathcal{X}_s$) satisfying constraints, unsafe set ($\mathcal{X}_u$) and state space ($\mathcal{X}$). $\mathcal{X}_f$ denotes the largest feasible safe set. Thus, in such a scenario, system trajectories ($T_r$) must remain in the safe set $\mathcal{X}_s$ and furthermore, avoid exploration in $\mathcal{X}_u$.}
\label{fig:Safe_set}
\end{figure}

A broad definition of safety is taken into account for a dynamical system that is being regulated by a control policy. \em In order for the system to be considered safe, it must be ensured that the system state trajectories will never enter any unsafe set under the existing control policy. By guaranteeing a safe set's forward invariance, safety can be ensured. That is, for every time $t$ ($t \geq 0$), all trajectories beginning in the set of safe states will remain there. Formally, given an unsafe set $\mathcal{X}_u \subseteq \mathcal{X}$ that does not contain the set of initial conditions $\mathcal{X}_0$, the system is safe if it never enters $\mathcal{X}_u$ when starting within $\mathcal{X}_0$. Safety is then defined for the system if it never enters the unsafe set $\mathcal{X}_u$ when starting within $\mathcal{X}_0$. \em The notion of safety can be further extended in terms of stability in the Lyapunov sense, which will be discussed in Section~\ref{subsection:CLF}. Fig.~\ref{fig:Safe_set} provides a notion of safe set.

\subsection{Constrained Markov Decision Process (CMDP) and Safe RL}\label{subsection:CMDP}
A Markov Decision Process (MDP) can be denoted by the tuple $<\mathcal{S}, \mathcal{A}, \mathcal{R},\mathbb{P},\mu, \gamma>$ ~\cite{meyn2022control}, where $\mathcal{S}$ and $\mathcal{A}$ denote the set of states and actions, respectively. $\mathcal{R}:\mathcal{S} \times \mathcal{A} \times \mathcal{S} \mapsto \mathbb{R}$ denotes the reward function, $\mathbb{P}:\mathcal{S} \times \mathcal{A} \times \mathcal{S} \mapsto [0,1]$ denotes the transition probability function, $\mu: \mathcal{S} \mapsto [0,1]$ is the initial probability distribution and $\gamma$ denotes the discount factor for future rewards. A policy $\pi: \mathcal{S} \mapsto \mathcal{P}(A)$ is a mapping from states to a probability distribution over actions and $\pi(a_t\vert s_t)$ is the probability of taking action $a$ under state $s$ at time $t$. Reinforcement learning (RL) involves an agent iteratively interacting with its environment, with the goal of learning a policy to maximize a given reward criterion. The goal of RL is to learn a policy $\pi$ that maximizes the discounted cumulative reward:
\begin{equation}
	\begin{aligned}
		\pi^* \in \text{arg} \max_{\pi} J_{\mathcal{R}}^{\pi} = \EX_{\tau \sim \pi} [\sum_{t=0}^{\infty} \gamma^t \mathcal{R}(s_t, a_t, s_{t+1})]
	\end{aligned}
\end{equation}
where $\tau$ denotes a state-action trajectory, and $\tau \sim \pi$ denotes trajectories sampled from the policy $\pi$. A CMDP is an extension of a MDP with an additional constraint set of cost functions $\mathcal{C}$~\cite{altman2021constrained}. The augmented tuple for a CMDP is then denoted by $<\mathcal{S}, \mathcal{A}, \mathcal{R},\mathcal{C},\mathbb{P},\mu, \gamma>$. The cost functions $\mathcal{C}_i \in \mathcal{C}$, $\mathcal{C}_i:\mathcal{S} \times \mathcal{A} \times \mathcal{S} \mapsto \mathbb{R}$ are constraint-associated cost functions. We call an action $a$ feasible if $a \in \mathcal{A}$ satisfies all of its constraints. The objective for safe RL is then to find a policy to maximize the long-term rewards while satisfying the constraints, denoted as:
\begin{equation}
\begin{aligned}
\pi^* \in \argmax_{\pi} \quad & J_{\mathcal{R}}^{\pi}=\EX_{\pi}\!\left[\sum_{t=0}^{\infty}\gamma^{t}\mathcal{R}(s_t,a_t,s_{t+1})\right] \\
\text{s.t.}\quad & J_{\mathcal{C}_i}^{\pi}=\EX_{\pi}\!\left[\sum_{t=0}^{\infty}\gamma^{t}\mathcal{C}_i(s_t,a_t,s_{t+1})\right]\le d_i,\quad i=1,\ldots,N_c .
\end{aligned}
\label{eqn:cmdp_expected_constraints}
\end{equation}

Equation~\eqref{eqn:cmdp_expected_constraints} captures expected cumulative CMDP constraints. This is weaker than pointwise hard safety, where one requires $g_i(s_t,a_t)\le0$ for all time steps, and different from chance-constrained safety, where one requires $\mathbb{P}_{\pi}(s_t\in\mathcal{S}_{\rm safe},\forall t)\ge1-\delta$. This distinction is important because many CMDP algorithms constrain expected cost but do not by themselves guarantee trajectory-wise forward invariance.

\begin{figure}[!t]
\centering
\resizebox{\columnwidth}{!}{%
\begin{tikzpicture}[
    >=Latex,
    block/.style={
        draw,
        rounded corners=6pt,
        thick,
        minimum width=3.7cm,
        minimum height=1.15cm,
        align=center,
        inner sep=6pt,
        font=\normalsize
    },
    arrow/.style={
        ->,
        thick,
        line width=0.9pt
    },
    lab/.style={
        font=\normalsize,
        fill=white,
        inner sep=1.8pt,
        text=black,
        align=center
    }
]

% -------------------------------------------------
% Nodes
% -------------------------------------------------
\node[block, fill=red!10]    (cert)   at (0,4.35)
    {Certificate layer\\CLF / CBF / CLBF};

\node[block, fill=blue!10]   (actor)  at (-4.8,0.95)
    {Actor / policy\\$\pi_\theta(a\vert s)$};

\node[block, fill=gray!10]   (filter) at (0,0.95)
    {Projection or\\QP safety filter};

\node[block, fill=orange!12] (critic) at (4.8,0.95)
    {Critic\\$V_\psi,\ Q_\psi$};

\node[block, fill=green!12]  (env)    at (0,-2.05)
    {Environment\\$s_t,\ r_t,\ c_t,\ s_{t+1}$};

% -------------------------------------------------
% Main actor--environment loop
% -------------------------------------------------
\draw[arrow]
    (env.west)
    .. controls (-2.8,-1.9) and (-4.5,-0.65) ..
    node[lab, left, pos=0.58] {$s_t$}
    (actor.south);

\draw[arrow]
    (actor.east) --
    node[lab, above, midway] {$a_t^{\rm RL}$}
    (filter.west);

\draw[arrow]
    (filter.south) --
    node[lab, right, midway] {$u_t$}
    (env.north);

\draw[arrow]
    (env.east)
    .. controls (2.8,-1.9) and (4.5,-0.65) ..
    node[lab, right, pos=0.58] {TD data}
    (critic.south);

% -------------------------------------------------
% Critic-to-actor learning signal
% Raised much higher to avoid overlap with constraints
% -------------------------------------------------
\draw[arrow]
    (critic.north)
    .. controls (3.4,3.15) and (-3.4,3.15) ..
    node[lab, above, pos=0.50] {Advantage / TD error}
    (actor.north);

% -------------------------------------------------
% Certificate connections
% -------------------------------------------------
% Constraints arrow: keep vertical, but move label to the LEFT
\draw[arrow]
    (cert.south) --
    node[lab, left, xshift=-2pt, pos=0.65] {constraints}
    (filter.north);

% To critic: route outward on right
\draw[arrow]
    (cert.east)
    .. controls (2.8,4.1) and (4.9,2.75) ..
    node[lab, right, pos=0.56] {loss terms}
    (critic.north);

% To actor: route outward on left
\draw[arrow]
    (cert.west)
    .. controls (-2.8,4.1) and (-4.9,2.75) ..
    node[lab, left, pos=0.56] {policy regularization}
    (actor.north);

\end{tikzpicture}%
}
\caption{Actor--critic safe RL pipeline showing where Lyapunov/barrier certificates can enter: as actor or critic losses, reward or value shaping, constrained policy updates, policy regularization, or a projection/QP safety filter that modifies the nominal RL action before execution.}
\label{fig:actor-critic}
\end{figure}
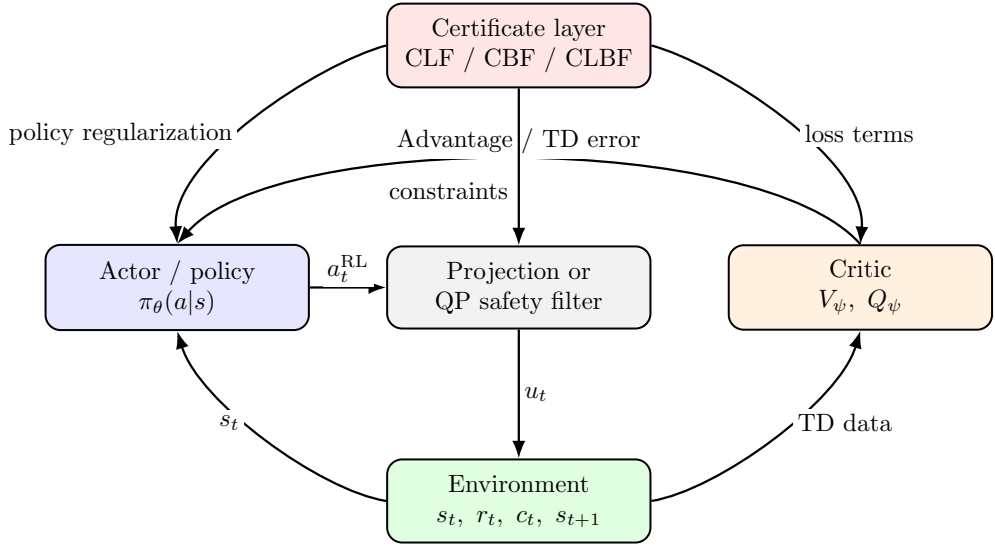

An excellent taxonomy for types of constraints for CMDP can be found in~\cite{liu2021policy}. A few key equations for the state-value function (\ref{eqn:Vfunc}), state-action-value or Q function (\ref{eqn:Qfunc}) and advantage function (\ref{eqn:Afunc}) are given below:
\begin{align}
	V_{\pi}(s) = \EX_{\pi}[\sum_{t=0}^{\infty} \gamma^t r_{t+1}\vert s_0=s] \label{eqn:Vfunc}\\ 
	Q_{\pi}(s,a) = \EX_{\pi}[\sum_{t=0}^{\infty} \gamma^t r_{t+1}\vert s_0=s, a_0=a] \label{eqn:Qfunc}\\
	A_{\pi}(s,a) = Q_{\pi}(s,a)- V_{\pi}(s) \label{eqn:Afunc}
\end{align}
Function approximators in RL are generally used to approximate the Value or Q function and/or the policy $\pi$ depending on the algorithm employed. Fig. \ref{fig:actor-critic} provides an overview of the actor-critic algorithm.

\subsection{Lyapunov Functions }\label{subsection:CLF}
\begin{figure}[t]
\centering
\resizebox{0.98\linewidth}{!}{%
\begin{tikzpicture}[
    >=Latex,
    font=\sffamily\small,
    panel/.style={draw=black!55, rounded corners=6pt, fill=gray!3, line width=0.7pt},
    heading/.style={font=\sffamily\bfseries\large, align=center},
    subhead/.style={font=\sffamily\footnotesize, text=black!62, align=center},
    eqbox/.style={draw=black!15, fill=white, rounded corners=3pt, inner sep=4pt, font=\sffamily\scriptsize, align=center, text width=3.6cm},
    cert/.style={font=\sffamily\scriptsize, align=center, text=black!72},
    traj/.style={-{Latex[length=2.0mm,width=1.4mm]}, line width=1.05pt, draw=orange!90!black},
    level/.style={draw=blue!55!black, line width=0.55pt},
    cbf/.style={draw=green!45!black, line width=1.05pt}
]

% =====================================================
% Left panel: Lyapunov
% =====================================================
\draw[panel] (-5.10,-2.55) rectangle (-0.30,2.55);
\node[heading] at (-2.70,2.18) {Lyapunov Function};
\node[subhead] at (-2.70,1.84) {asymptotic stability};
\node[eqbox] at (-2.70,1.23)
{$V(x^\star)=0,\;V(x)>0$\\[-1pt]
$\dot V(x)<0$ for $x\ne x^\star$};

% Level sets
\draw[level, fill=blue!5]  (-2.70,-0.55) ellipse (1.55 and 1.00);
\draw[level, fill=blue!10] (-2.70,-0.55) ellipse (1.07 and 0.69);
\draw[level, fill=blue!18] (-2.70,-0.55) ellipse (0.60 and 0.38);

% Equilibrium and trajectory
\fill[red!80!black] (-2.70,-0.55) circle (2.0pt);
\node[font=\sffamily\scriptsize, anchor=west] at (-2.58,-0.55) {$x^\star$};
\draw[traj, smooth, samples=200, domain=0:7.2, variable=\t]
    plot ({-2.70 + 1.37*exp(-0.23*\t)*cos(172*\t)},
          {-0.55 + 0.88*exp(-0.23*\t)*sin(172*\t)});
\node[cert] at (-2.70,-2.18) {decreasing level sets $V=c$};

% =====================================================
% Right panel: Barrier
% =====================================================
\draw[panel] (0.30,-2.55) rectangle (5.10,2.55);
\node[heading] at (2.70,2.18) {Barrier Function};
\node[subhead] at (2.70,1.84) {forward invariance};
\node[eqbox] at (2.70,1.23)
{$\mathcal C=\{x:h(x)\le0\}$\\[-1pt]
$\dot h(x)\le-\alpha(h(x))$};

% Safe set
\draw[cbf, fill=green!9] (2.70,-0.55) ellipse (1.55 and 1.00);
\node[font=\sffamily\footnotesize, text=green!35!black] at (3.58,0.18) {$\mathcal C$};
\node[font=\sffamily\scriptsize, text=green!35!black, anchor=west] at (4.02,-0.55) {$\partial\mathcal C$};

% Invariant trajectory
\draw[traj]
    (1.78,-0.55)
    .. controls (1.91,0.05) and (2.52,0.30) .. (3.05,0.08)
    .. controls (3.62,-0.18) and (3.47,-0.92) .. (2.78,-1.04)
    .. controls (2.05,-1.16) and (1.67,-0.73) .. (1.98,-0.24);
\fill[red!80!black] (1.78,-0.55) circle (2.0pt);
\node[font=\sffamily\scriptsize, anchor=east] at (1.60,-0.55) {$x_0$};

% Boundary arrows indicating inward condition
\draw[-{Latex[length=1.6mm,width=1.1mm]}, draw=green!45!black, line width=0.75pt] (2.70,0.45) -- (2.70,0.22);
\draw[-{Latex[length=1.6mm,width=1.1mm]}, draw=green!45!black, line width=0.75pt] (1.34,-0.55) -- (1.58,-0.55);
\draw[-{Latex[length=1.6mm,width=1.1mm]}, draw=green!45!black, line width=0.75pt] (4.06,-0.55) -- (3.82,-0.55);
\node[cert] at (2.70,-2.18) {trajectories remain in $\mathcal C$};

\end{tikzpicture}%
}
\caption{Lyapunov and barrier functions provide complementary safety certificates for control and reinforcement learning. \emph{Left:} A Lyapunov function certifies stability by ensuring that trajectories move toward lower level sets and converge to the equilibrium $x^\star$. \emph{Right:} A barrier function certifies forward invariance of the safe set $\mathcal C$ by enforcing a boundary condition that prevents trajectories initialized in $\mathcal C$ from leaving the constraint-admissible region.}
\label{fig:Lyapunov_barrier}
\end{figure}
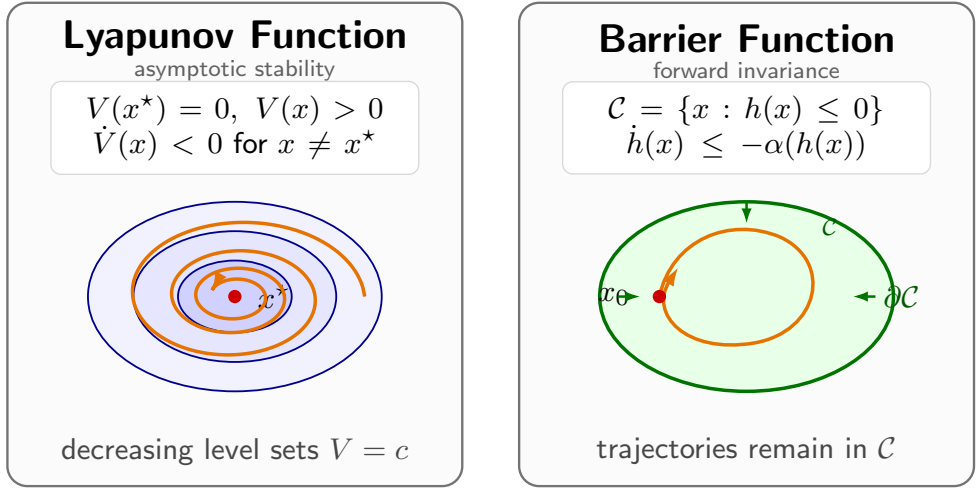

Fig.~\ref{fig:Lyapunov_barrier} provides an insight into utility of Lyapunov and barrier functions in control theory and RL.
A continuously differentiable function $V:\mathcal{X} \mapsto \mathbb{R}$ is a Lyapunov function if~\cite{khalil2002nonlinear}:
\begin{subequations}
	\begin{gather}
		V(x_d) = 0 \\
		V(x) > 0, \quad \forall x \in \mathcal{X} \backslash \{x_d\} \\
		\nabla V(x)f_{cl}(x) \leq 0, \quad x \in \mathcal{X}
	\end{gather}
\end{subequations}
where $x_d \in \mathcal{X}$ is the desired state for the system and $f_{cl}$ is the closed loop dynamics of the system. $\nabla V(x)f_{cl}(x)$ denotes the Lie derivative of $V$ along the function $f_{cl}$. If only $\dot V(x)\le0$ is available, the condition establishes non-increase of the Lyapunov function and typically requires an additional LaSalle-type invariance argument to conclude asymptotic convergence. Control Lyapunov functions provide formal guarantees for stabilizability of open-loop systems, which implies the existence of a controller that drives the closed loop system to stability. Formally, a control Lyapunov function (CLF) is a criterion that provides asymptotically stabilizing controllers for a general nonlinear dynamical system, which was first generalized by Sontag~\cite{sontag1989universal}. For a control affine form in Section \ref{subsection:SafetyCriterion}, a CLF $V$ is a smooth (continuous and differentiable up to some order), proper ($V(x) \rightarrow \infty$ as $\Vert x \Vert \rightarrow \infty$) and positive definite ($V(x) > 0$ and $V(0) = 0$ for $x\neq 0$) function:
\begin{equation}
	\begin{aligned}
		V:\mathbb{R}^n \mapsto \mathbb{R}
	\end{aligned} 	
\end{equation}
then V certifies asymptotic stabilizability about $x_d$ if,
\begin{subequations}
	\begin{gather}
		V(x_d) = 0 \\
		V(x) > 0, \quad \forall x \in \mathcal{X} \backslash \{x_d\} \\
		\inf_{u\in \mathcal{U}} [L_fV(x) + L_gV(x)u] \leq 0, \quad \forall x \in \mathcal{X}
	\end{gather}
\end{subequations}
where $L_f$ and $L_g$ are Lie derivatives along $f$ and $g$, respectively. The control policy can then be solved by formulating these conditions as a Quadratic Programming (QP) problem (because affine in $u$) as follows:
\begin{equation}
\begin{aligned}
(u^\star,\delta^\star)=\arg\min_{u\in\mathcal{U},\delta\ge0}\quad & \Vert u-u_{\rm nom}\Vert^2+p\delta^2 \\
\text{s.t.}\quad & L_fV(x)+L_gV(x)u \le -\alpha_V(V(x))+\delta,
\end{aligned}
\label{eqn:CLF_QP_revised}
\end{equation}
Here $u_{\rm nom}$ may be the action proposed by the RL policy, a model-based controller, or a human/reference command. The slack variable $\delta$ is often included to maintain feasibility when CLF objectives conflict with other constraints.

\subsection{Barrier Functions }\label{subsection:CBF}

Barrier functions were first introduced in optimization literature, which are added to cost functions to avoid undesirable regions. CBFs have proven to be effective in defining safe sets in the control community~\cite{ames2019control}. Approximating system models helps alleviate the need for complete prior system knowledge which is required for the formulation of CBFs. We briefly cover theoretical aspects of control barrier functions in this review, a comprehensive review of theory and applications of control barrier functions can be found in~\cite{ames2019control} and barrier functions in~\cite{polyak1992modified}. Consider a compact set $\mathcal{C}$, defined as the zero sublevel set of a barrier function $h:\mathcal{X}\mapsto \mathbb{R}$ ($\mathcal{C} = {x:h(x)\leq 0}$). We use a sublevel-set convention for $h$; under the more common superlevel-set convention $\mathcal{C}=\{x:h(x)\ge0\}$, the signs in the CBF inequalities are reversed. From Proposition 1 in \cite{ames2016control}, if there exists a strictly increasing scalar function $\alpha:\mathbb{R}\mapsto\mathbb{R}$ such that $\alpha(0)=0$ (an extended class-$\mathcal{K}$ function) and
\begin{equation}
	\frac{dh}{dt} \leq -\alpha(h(x)), \quad \forall x \in \mathcal{X}
\end{equation}
then $h$ is a barrier function and $\mathcal{C}$ is forward invariant for the closed loop system $\dot{x}=f_{cl}(x)$. A CBF can be defined as:
\begin{equation}
	\begin{gathered}
		\inf_{u\in \mathcal{U}} [L_fh(x) + L_gh(x)u + \alpha(h(x))] \leq 0, \quad \forall x \in \mathcal{X}
	\end{gathered}
\end{equation}
Similar to CLF QP formulation, CBFs can also be formulated as a QP since the condition is affine in $u$.
\begin{equation}
	\begin{gathered}
		\min_{u \in \mathcal{U}} \Vert u \Vert^2 \\
		\text{s.t. } L_fh(x) + L_gh(x)u + \alpha(h(x))] \leq 0 \label{eqn:CBF_QP}
	\end{gathered}
\end{equation}

\subsection{Running Example: Cart-Pole Stabilization}\label{subsec:cartpole_example}
Most RL algorithms interact with a sampled environment, while CLF/CBF theory is often introduced in continuous time. For a discrete-time system $x_{k+1}=F(x_k,u_k)$, a Lyapunov decrease condition is commonly written as
\begin{equation}
V(x_{k+1})-V(x_k)\le -\alpha_V(\Vert x_k-x_d\Vert),
\label{eqn:dt_lyap}
\end{equation}
and a discrete-time barrier condition under a sublevel safe-set convention $\mathcal{C}=\{x:h(x)\leq0\}$ can be written as
\begin{equation}
h(x_{k+1})\le (1-\eta_b)h(x_k),\qquad \eta_b\in[0,1].
\label{eqn:dt_cbf}
\end{equation}
These discrete-time forms are useful for RL because they can be evaluated directly from sampled transitions $(s_k,a_k,s_{k+1})$ without explicitly computing Lie derivatives.

Throughout Sections~\ref{section:ControlLF}--\ref{section:CLFandCBF} we use the canonical Cart-Pole stabilization task as a running example to ground the abstract formulations in a concrete control problem. The state $x = [p,\dot p,\theta,\dot\theta]^\top \in \mathbb{R}^4$ comprises cart position, cart velocity, pole angle, and pole angular velocity; the control $u \in \mathbb{R}$ is the horizontal force on the cart. The system dynamics $\dot x = f(x) + g(x)u$ are control-affine and locally Lipschitz on the operating region. Two safety constraints are imposed: a track-position bound $|p| \leq p_{\max}$ and a pole-angle bound $|\theta| \leq \theta_{\max}$, encoded as barrier functions $h_1(x) = p^2 - p_{\max}^2 $ and $h_2(x) = \theta^2 - \theta_{\max}^2$ whose sublevel sets jointly define the safe set $\mathcal{C}$. A natural Lyapunov candidate is $V(x) = \tfrac{1}{2}x^\top P x$ with $P \succ 0$ derived from a linearization-based LQR; this candidate serves either as the critic in value-as-Lyapunov methods, as the CLF in CLF--CBF--QP filters, or as a regularizer in Lyapunov reward shaping. We refer back to this setup in callouts within each major method block.
%%%%%%%%%%%%%%%%%%%%%%%%%%%%%%%%%%%%%%%%%%%%%%%%%%%%%%%%%%%%%%%%%%%%%%%%%%%%%%%%%%%%%%%%%%%%%%%%%%%%%%%%%%%%%%%%%%%
%%%%%%%%%%%%%%%%%%%%%%%%%%%%%%%%%%%%%%%%%%%%%%%%%%%%%%%%%%%%%%%%%%%%%%%%%%%%%%%%%%%%%%%%%%%%%%%%%%%%%%%%%%%%%%%%%%%
\section{Lyapunov Functions for Reinforcement Learning}\label{section:ControlLF}
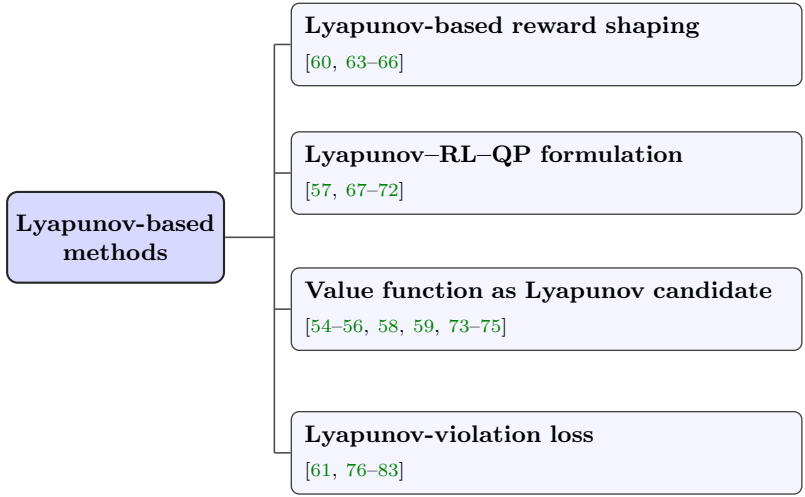
\begin{figure}[t]
\centering
\begin{tikzpicture}[
  >=stealth, line width=0.5pt,
  rootnode/.style={
    rectangle, rounded corners=4pt, draw=black!85, line width=0.8pt,
    fill=blue!15, minimum width=2.6cm, minimum height=1.2cm,
    align=center, font=\bfseries\small
  },
  catnode/.style={
    rectangle, rounded corners=3pt, draw=black!75, line width=0.5pt,
    fill=blue!4, text width=6.4cm, align=left, inner sep=5pt, font=\small
  },
  branch/.style={line width=0.6pt, draw=black!70}
]
  \node[rootnode] (root) at (0, 0) {Lyapunov-based\\methods};

  \node[catnode] (c1) at (5.7, 2.55) {%
    \textbf{Lyapunov-based reward shaping}\\[2pt]%
    \footnotesize \cite{chow2018lyapunov,chow2019lyapunov,dong2020principled,ugurlu2025lyapunov,xia2024estimating}};

  \node[catnode] (c2) at (5.7, 0.85) {%
    \textbf{Lyapunov--RL--QP formulation}\\[2pt]%
    \footnotesize \cite{kumar2017diagonal,huh2020safe,jeddi2021lyapunov,westenbroek2020learning,cao2023toward,tesfazgi2024stable,long2025certifying}};

  \node[catnode] (c3) at (5.7, -0.95) {%
    \textbf{Value function as Lyapunov candidate}\\[2pt]%
    \footnotesize \cite{vamvoudakis2015asymptotically,kamalapurkar2016model,berkenkamp2017safe,chang2021stabilizing,westenbroek2022lyapunov,lopez2024decomposing,yao2025lyapunov,tan2025finite}};

  \node[catnode] (c4) at (5.7, -2.85) {%
    \textbf{Lyapunov-violation loss}\\[2pt]%
    \footnotesize \cite{10155918,krishna2023finite,kumar2023task,wang2024actor,KUSHWAHA202484,mccutcheon2025neural,hao2024lyapunov,chen2026safe,kang2022lyapunov}};

  \coordinate (j) at (2.1, 0);
  \draw[branch] (root.east) -- (j);
  \draw[branch] (j |- c1.west) -- (j |- c4.west);
  \draw[branch] (j |- c1.west) -- (c1.west);
  \draw[branch] (j |- c2.west) -- (c2.west);
  \draw[branch] (j |- c3.west) -- (c3.west);
  \draw[branch] (j |- c4.west) -- (c4.west);
\end{tikzpicture}
\caption{Classification of methods using Lyapunov functions in safe RL. Categories correspond to how the Lyapunov structure enters the learning loop: as a reward-shaping term, as an explicit QP constraint, as the value/critic function itself, or as a violation-penalty loss.}
\label{fig:classification_Lyap}
\end{figure}

The closed-loop system's stability assurances can be used to ensure safety guarantees. These methods frequently rely on control Lyapunov functions (CLFs) for Lyapunov stability verification presented in Section~\ref{subsection:CLF}. CLFs were formalized for a general nonlinear dynamical system by Sontag~\cite{sontag1989universal}. An interesting property that a lot of work in safe RL leverages is the fact that level sets of CLFs are both attractive and invariant. Fig.~\ref{fig:classification_Lyap} shows a brief classification of using Lyapunov functions approaches in RL. The earliest work using Lyapunov functions with RL was proposed by Perkins et al.~\cite{perkins2002lyapunov}, where the authors use Lyapunov functions to design multiple controllers which the control policy can switch between, which is designed using RL (agent decides when to switch between Lyapunov-based control laws). The system is formulated as a MDP and the authors implement state-action-reward-state-action (SARSA) ($\lambda$) RL algorithm with cerebellar model articulation controller (CMAC) function approximators~\cite{lane1992theory} for action-value functions. The use of Lyapunov function ensures safety guarantees for any switching policy and furthermore, empirically showed accelerated learning.

\emph{Cart-Pole instantiation.} On the running example of Section~\ref{subsec:cartpole_example}, value-as-Lyapunov methods set the critic equal to the LQR-derived $V(x) = \tfrac{1}{2}x^\top P x$ near the upright equilibrium and grow it through actor--critic updates; reward-shaping methods (Section~\ref{section:ControlLF} below) instead add $\lambda(\gamma V(x') - V(x))$ to the reward to penalize Lyapunov increase along trajectories.

In~\cite{vamvoudakis2015asymptotically}, the authors propose an actor-critic RL algorithm (with NNs as function approximators) and a Lyapunov function as the optimal-value function for a deterministic nonlinear dynamical system. The authors prove asymptotic stability of the proposed algorithm (Section IV in the work). Similarly, in~\cite{kamalapurkar2016model}, the authors consider a Lyapunov candidate function as the value function and use NNs to approximate the solution for the Hamilton-Jacobi-Bellman (HJB) equation. The authors develop a concurrent learning-based implementation of RL and use Lyapunov analysis to guarantee stability for an approximate dynamic programming (ADP) problem. Similar to the previous work, authors in ~\cite{lopez2024decomposing} compute a control Lyapunov-value function (CLVF) using Hamilton-Jacobi reachability (HJR) analysis. HJR analysis computes a reach--avoid set which can be broadly described as set of states from which the system can driven using a control input to a target set while satisfying operational constraints. In~\cite{kumar2017diagonal}, the authors use ADP for a nonlinear dynamical system with weight updates using the Lyapunov stability criterion. NNs are used as function approximator for the actor, critic networks and model approximation. The Lyapunov stability criterion based weight updates perform better than gradient descent for parameter variation and disturbance signal effects. Thus, in \cite{vamvoudakis2015asymptotically,kamalapurkar2016model, kumar2017diagonal} a similar approach of considering the value function as the candidate Lyapunov function is considered for an ADP formulation, with \cite{kumar2017diagonal} only differing in the parameter update criterion. Authors in \cite{yao2025lyapunov} treat the Q-function as a candidate Lyapunov function and present a distributed multi-agent architecture.

Berkenkamp et al.~\cite{berkenkamp2017safe} propose the use of Gaussian Process (GP) to approximate the system model online and uses the region of attraction (states inside region of attraction are safe) to formulate the safety constraints. The region of attraction is selected by taking the largest level set of Lyapunov function. Thus, the policy optimization problem has a constraint in terms of the Lyapunov decrease condition which is given as follows:
\begin{subequations}
	\begin{gather}
		L^{dec}_n = \{(x,u)\in \mathcal{X}_{\tau} \times \mathcal{U} \lvert p_n(x,u) - V(x) < - L_{\Delta v}\tau \} \\
		L_{\Delta v} := L_vL_f(L_{\pi}+1)+L_v	
	\end{gather}
\end{subequations}
where, $p_n{x,u}$ is the upper bound on the $V(f(x,u))$, $V(\cdot)$ is the Lyapunov function and dynamics of system is given by $f(x,u)$. $\tau$ is the discretization constant for the discrete dynamics and $\mathcal{X}_{\tau}$ is the discretization of $\mathcal{X}$. $L_v$, $L_f$ and $L_{\pi}$ are the Lipschitz constants for the Lyapunov function, dynamics and the control policy set $\Pi_L$ (Assumption $1$ in the cited work), respectively. Thus, largest level set of $v$ for which all state-action pairs that correspond to discrete states within $\mathcal{X}_{\tau}$ are contained in the set $L^{dec}_n$. \cite{xia2024estimating} follows a similar approach as well with the difference being incorporation of the empirical Lyapunov risk term in the reward function. A drawback of these approaches is that the initial policy is assumed to be known and safe for the system. Expected violation of Lyapunov conditions is used as the loss function for critic in \cite{wang2024actor}. The authors propose a novel physics informed actor-critic architecture where a Zubov function \cite{zubov1964methods} (maximal Lyapunov function) is employed to approximate the critic function which denotes the physics informed part in the algorithm. Overall, the Zubov function (critic) and controller (actor) are iteratively updated ensuring the proposed method satisfies the Lyapunov conditions.

Compared to most prior work based on control affine nonlinear dynamical systems, in~\cite{chow2018lyapunov}, authors formulate the safe RL problem as a CMDP. The safety constraints in the CMDP are modeled as constraint cost function (upper bound on cost) using Lyapunov functions, thereby reformulating the value function as a cost-value function. The following problem is solved in this case: Given an initial condition $x_0$ and a upper-bound on the expected cumulative constraint cost $d_0$, solve
\begin{equation}
	\min_{\pi\in\Delta} \{\mathcal{C}_{\pi}(x_0): \mathcal{D}_{\pi}(x_0) \leq d_0\}
\end{equation}
If there is a non-empty solution, the optimal policy is denoted by $\pi^{*}$. $\Delta$ denotes the set of Markov stationary policies, $\mathcal{C}_{\pi}$ is the cost function and $\mathcal{D}_{\pi}$ is the safety constraint function. Thus, the goal is to create a Lyapunov function $v(x)$ such that,
\begin{equation}
	v(x) \geq T_{\pi^{*},d}[v](x), \quad v(x_0) \leq d_0 \label{eqn:bellman}
\end{equation}
where, $T_{\pi^*,d}$ is the Bellman operator with respect to optimal policy $\pi^*$ and cost constraint function $d$. 
This framework is extended to DP and RL algorithms (safe policy iteration, safe value iteration, safe Q-learning etc.) in this work. Chow et al.~\cite{chow2019lyapunov} also propose safe policy optimization for continuous action CMDP formulations and test it on MuJoCo and real-world indoor navigation tasks. Two approaches: $\theta$-projection (constrained optimization which reformulates policy gradient (PG) with a projection of the policy parameters onto a feasible set using Lyapunov functions) and $a$-projection (projection of unconstrained action on the Lyapunov hyperplane) are proposed for solving on and off policy gradient algorithms. Another instance of using CMDP formulation is proposed by authors in \cite{cao2023toward}, where the Lyapunov drift term is used as a constraint for QP problem that projects the RL action to a safe set. This approach is similar to the CLF-QP formulation with CLF constraints replaced by Lyapunov drift terms. 

\begin{sidewaystable}[htpb]
\centering
\caption{Summary of safe RL approaches using Lyapunov functions.}
\label{tab:CLF}
\small
\setlength{\tabcolsep}{4pt}
\renewcommand{\arraystretch}{0.95}
\begin{tabular}{@{}p{0.20\linewidth} p{0.40\linewidth} p{0.18\linewidth} p{0.20\linewidth}@{}}
\toprule
\textbf{Reference} & \textbf{Method} & \textbf{RL algorithm} & \textbf{Simulation} \\
\midrule

Perkins \& Barto (2002) \cite{perkins2002lyapunov} & Switching between Lyapunov control laws & SARSA($\lambda$) & Pendulum \\
Vamvoudakis et al.\ (2015) \cite{vamvoudakis2015asymptotically} & Lyapunov candidate as value function & ADP & Van der Pol, power plant \\
Kamalapurkar et al.\ (2016) \cite{kamalapurkar2016model} & Lyapunov candidate as value function & ADP & LQT problem \\
Berkenkamp et al.\ (2017) \cite{berkenkamp2017safe} & ROA safe set via largest Lyapunov level set & ADP & Inverted pendulum \\
Kumar et al.\ (2017) \cite{kumar2017diagonal} & Lyapunov criterion-based weight updates & ADP & Nonlinear systems \\
Chow et al.\ (2018) \cite{chow2018lyapunov} & Lyapunov as CMDP cost constraints & Multiple DP/RL & 2D grid world \\
Chow et al.\ (2019) \cite{chow2019lyapunov} & $\theta$- and $a$-projection onto Lyapunov-safe set & Multiple PG & MuJoCo \\
Dong et al.\ (2020) \cite{dong2020principled} & Lyapunov-based reward shaping & DQN, PPO & OpenAI Gym, MuJoCo \\
Huh et al.\ (2020) \cite{huh2020safe} & Lyapunov-based policy optimization & Tabular-Q, DDPG & Double integrator, Reacher \\
Westenbroek et al.\ (2020) \cite{westenbroek2020learning} & CLF rate-of-dissipation constraint on policy & SAC & Double pendulum, bipedal walker \\
Jeddi et al.\ (2021) \cite{jeddi2021lyapunov} & Lyapunov constraints on policy & DQN, PPO, TRPO & Grid world \\
Chang et al.\ (2021) \cite{chang2021stabilizing} & Self-learned almost-Lyapunov critic & PPO & OpenAI Gym, MuJoCo \\
Westenbroek et al.\ (2022) \cite{westenbroek2022lyapunov} & CLF as value function under low discount & SAC & Cartpole, A1 quadruped \\
Hejase et al.\ (2023) \cite{10155918} & Joint CLF + dynamics DNNs; violation penalty & DDPG & highway-env \\
Krishna et al.\ (2023) \cite{krishna2023finite} & Controlled finite-time Lyapunov exponent analysis & MPC, DDPG & Unsteady fluid flow \\
Kumar et al.\ (2023) \cite{kumar2023task} & Lyapunov drift/penalty in policy loss & Multi-agent DDPG & PTV Vissim (traffic) \\
Cao et al.\ (2023) \cite{cao2023toward} & Lyapunov safe set with action projection & PPO & Custom \\
Lopez et al.\ (2024) \cite{lopez2024decomposing} & Subsystem-decomposed CLF via HJ reachability & SAC, PPO & Dubins car, lunar lander, drone \\
Wang et al.\ (2024) \cite{wang2024actor} & Zubov critic; minimize Lyapunov violation & Actor--critic & Inverted pendulum, Van der Pol \\
Tesfazgi et al.\ (2024) \cite{tesfazgi2024stable} & CLF learning from demonstrations & Inverse RL & LASA dataset \\
Kushwaha et al.\ (2024) \cite{KUSHWAHA202484} & Lyapunov-RL with Koopman-identified dynamics & DDPG & Vehicle parking \\
Yao et al.\ (2025) \cite{yao2025lyapunov} & State-value as Lyapunov; distributed actor--critic & Actor--critic & MATLAB \\
McCutcheon et al.\ (2025) \cite{mccutcheon2025neural} & Neural Lyapunov + Lyapunov-risk loss & SAC & Gymnasium \\
Ugurlu et al.\ (2025) \cite{ugurlu2025lyapunov} & Lyapunov-based penalty in reward & PPO & OpenAI Gym, PX4 quadrotor \\
Hao et al.\ (2025) \cite{hao2024lyapunov} & Lyapunov constraints on policy loss & Multiple RL/SRL & Microgrid (custom) \\
Xia et al.\ (2025) \cite{xia2024estimating} & Lyapunov ROA reward term & Model-based RL & USV (custom) \\
Long et al.\ (2025) \cite{long2025certifying} & Lyapunov certificates as constraints + penalty & PPO, SAC, TD-MPC & Gymnasium, DeepMind Control \\

\bottomrule
\end{tabular}
\end{sidewaystable}

Reward shaping is another method for accelerating the learning process in RL while preserving optimality of the policy~\cite{ng1999policy}. Authors in~\cite{dong2020principled, ugurlu2025lyapunov} explore reward shaping using Lyapunov stability theory. A potential-based Lyapunov shaping term can be written as,
\begin{equation}
r'(s,a,s') = r(s,a,s') + \lambda\big(\gamma_{\rm RL}\Phi(s')-\Phi(s)\big), \qquad \Phi(s)=-V(x(s)),
\label{eqn:reward_lyap}
\end{equation}
where $\lambda$ adjusts the influence of the shaping term and $\Phi$ is chosen so that reducing the Lyapunov certificate increases the shaped reward. For cart-pole, this amounts to rewarding actions that reduce the quadratic stabilization certificate between consecutive sampled states. An asymptotically unbiased optimal greedy policy is also introduced by maximizing Q-function combined with an additional term similar to (\ref{eqn:reward_lyap}). Deep Q-learning networks (DQN) and proximal policy optimization (PPO) are used to verify the results in MuJoCo and OpenAI gym. Huh et al.~\cite{huh2020safe} propose the use of Lyapunov-based shielding to compute a safe policy and also an efficient safe exploration using Lyapunov constraints for a MDP formulation. Safe policies are sampled from the Lyapunov induced safe set which satisfies the probabilistic safety condition up to a threshold similar to the approach using (\ref{eqn:bellman}). Efficiency in exploration is ensured using experience replay and for safe exploration the authors propose an exploratory policy (different from the safe policy) which drives an agent around the boundary of the safe set. The methodology is tested for tabular Q-learning and deep deterministic policy gradient (DDPG) algorithms for a double integrator and robotic simulations. In~\cite{jeddi2021lyapunov}, the authors propose a similar Lyapunov-based safe policy formulation and a Transformer Neural Network-based approach to account for uncertainty in a CMDP formulation. The Transformer model is used as a memory buffer to capture long-term dependencies for uncertainty, which is used to predict the mean and variance of violating a safety constraint over a horizon. The mean and variance as fed as inputs to the risk averse action selection method which minimize the joint cost of selecting an action to minimize the mean and variance of violating the safety constraint over a horizon. The proposed algorithm is tested in 2D grid world with static and dynamic obstacles using DQN, PPO and trust region policy optimization (TRPO) algorithms.

A self-learned Lyapunov critic is a function that estimates the region of attraction for a closed loop controlled system. This is achieved by estimating the empirical Lyapunov risk, which provides a measure of how likely the system is to diverge from a given state. Almost Lyapunov conditions~\cite{liu2020almost} solve the problem of incomplete model information by relaxing the Lyapunov conditions, i.e., a set $\Omega$ is introduced which contains the states that violate the Lyapunov conditions, however, as long as each component of $\Omega$ is small enough (bounded) the violations do not affect stability. Chang et al.~\cite{chang2021stabilizing} leverage these properties to propose self-learned almost Lyapunov critics for policy optimization. Authors introduce use of empirical Lyapunov risk ($R_{f,N,\rho}(\theta)$), which acts as the loss function for Lyapunov critic NN and is given as follows:
\begin{equation}
	\frac{1}{N} \sum_{i=1}^{N}(\max(-V_{\theta}(x_i),0)+\max(0,L_fV_{\theta}(x_i))) + V^2_{\theta}(0)
\end{equation}
where, $x_1,\dots,x_N$ are states sampled with respect to distribution $\rho$, $V_{\theta}$ is the parameterized Lyapunov function (NNs) and $L_fV_{\theta}$ is the Lie derivative of the Lyapunov function along the dynamics $f$. The key challenge in this approach is approximating the Lie derivative of the Lyapunov function along $f$ in a model-free setting. Finite differences is used by the authors in this approach to approximate $L_fV_{\theta}$ as follows:
\begin{equation}
	L_{f,\Delta t}V(x) = \frac{1}{\Delta t}(V(x') - V(x))
\end{equation}
Stochastic gradient descent is used to minimize the loss function. The proposed algorithm, ``Policy optimization with self-learned almost Lyapunov critics algorithm'', follows a structure similar to PPO with the addition of Lyapunov risk minimization for the parameter $\theta$ updates. The algorithm's performance is compared with Lyapunov actor-critic, soft actor critic and PPO for inverted pendulum, quadrotor control, automobile path-tracking and MuJoCo Hopper. A similar approach is proposed in \cite{mccutcheon2025neural, hao2024lyapunov}. Westenbroek et al.~\cite{westenbroek2020learning} propose the use of a CLF for an inaccurate model of the system and use the rate of dissipation of the CLF as a constraint (threshold for minimum acceptable rate) on closed-loop controller. The controller parameters are optimized using model-free policy optimization using data sampled from the system. The framework is tested on a double pendulum and under-actuated bipedal robots using soft actor-critic (SAC) algorithm. Westenbroek et al.~\cite{westenbroek2022lyapunov} also propose the use of CLF as a value function and show that smaller values for discount factor lead to stabilizing controllers with the choice of CLF as the value function. The proposed algorithm is tested on cartpole and an A1 quadruped using the SAC algorithm with augmented value function approach. Authors in~\cite{10155918} train a DDPG agent based on the violations of Lyapunov conditions for nonlinear vehicle dynamics model in a lane-following environment. The two-step approach first learns system dynamics and a CLF using NNs for a current policy and secondly, trains a DDPG agent to maximize reward and minimize violation of Lyapunov conditions using the learned CLF. A similar approach is followed in \cite{KUSHWAHA202484} with changes to using Koopman operator theory to identify the system dynamics and avoid training an extra DNN. 

A recent direction relaxes the requirement of an exact Lyapunov decrease on the full closed-loop dynamics. Chen et al.~\cite{chen2026safe} apply this idea to aerospace control via \emph{model-relaxed Lyapunov stability}: a nominal Lyapunov candidate is derived from a simplified rigid-body model, and the residual between simplified and true dynamics is bounded online via concentration inequalities, with safe RL operating on the residual. This allows certificate validity under realistic aerodynamic uncertainties and bridges the gap between toy-problem certificates and high-dimensional safety-critical aerospace applications.

Multi-agent RL (MARL) has seen significant theoretical contribution over the last decade, given its wide range of real-world applications. Authors in~\cite{kumar2023task} propose a Lyapunov-based multi-agent DDPG (L-MADDPG) for task scheduling and resource allocation of vehicle edge computing (VEC) assisted vehicle networks. The loss function in MADDPG is augmented with the Lyapunov drift and penalty terms.     

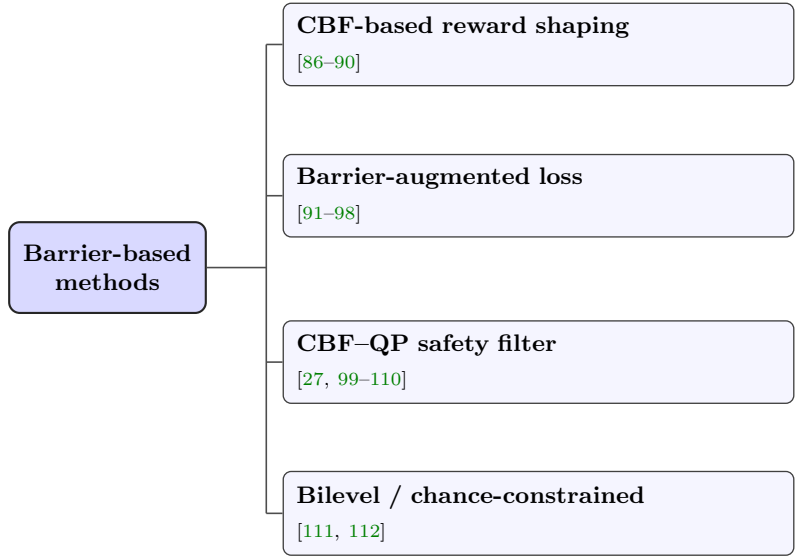
\begin{figure}[t]
\centering
\begin{tikzpicture}[
  >=stealth, line width=0.5pt,
  rootnode/.style={
    rectangle, rounded corners=4pt, draw=black!85, line width=0.8pt,
    fill=blue!15, minimum width=2.6cm, minimum height=1.2cm,
    align=center, font=\bfseries\small
  },
  catnode/.style={
    rectangle, rounded corners=3pt, draw=black!75, line width=0.5pt,
    fill=blue!4, text width=6.4cm, align=left, inner sep=5pt, font=\small
  },
  branch/.style={line width=0.6pt, draw=black!70}
]
  \node[rootnode] (root) at (0, 0) {Barrier-based\\methods};

  \node[catnode] (c1) at (5.7, 2.95) {%
    \textbf{CBF-based reward shaping}\\[2pt]%
    \footnotesize \cite{marvi2021safe,liu2023safe2,ranjan2024barrier,liu2025safe,xie2025certificated}};

  \node[catnode] (c2) at (5.7, 0.95) {%
    \textbf{Barrier-augmented loss}\\[2pt]%
    \footnotesize \cite{ohnishi2019barrier,zhang2022barrier,zhao2022safe,liu2023safe1,yang2023model,zhao2023barrier,zhang2024constrained,dey2024p2bpo}};

  \node[catnode] (c3) at (5.7, -1.25) {%
    \textbf{CBF--QP safety filter}\\[2pt]%
    \footnotesize \cite{cheng2019end,emam2021safe,ma2021model,cai2021safe,wang2022ensuring,huang2022barrier,song2022safe,brunke2022barrier,cheng2023safe,yang2024enhancing,dinh2024towards,hou2024safe,kalaria2024disturbance}};

  \node[catnode] (c4) at (5.7, -3.25) {%
    \textbf{Bilevel / chance-constrained}\\[2pt]%
    \footnotesize \cite{wang2023enforcing,sabouni2024reinforcement}};

  \coordinate (j) at (2.1, 0);
  \draw[branch] (root.east) -- (j);
  \draw[branch] (j |- c1.west) -- (j |- c4.west);
  \draw[branch] (j |- c1.west) -- (c1.west);
  \draw[branch] (j |- c2.west) -- (c2.west);
  \draw[branch] (j |- c3.west) -- (c3.west);
  \draw[branch] (j |- c4.west) -- (c4.west);
\end{tikzpicture}
\caption{Classification of methods using barrier functions in safe RL. Categories correspond to how the CBF condition enters the learning loop: as reward shaping, as an actor-loss penalty (with or without verification), as a per-step QP safety filter, or via bilevel/chance-constrained formulations.}
\label{fig:classification_barrier}
\end{figure}
%%%%%%%%%%%%%%%%%%%%%%%%%%%%%%%%%%%%%%%%%%%%%%%%%%%%%%%%%%%%%%%%%%%%%%%%%%%%%%%%%%%%%%%%%%%%%%%%%%%%%%%%%%%%%%%%%%%
\subsection{Classification and Discussion}
\label{subsec:lyap_discussion_open_problems}

Table \ref{tab:lyap_rl_classification} provides a tabular comparison of each methodology and its shortcomings. The body of work surveyed in Section~\ref{section:ControlLF} demonstrates that Lyapunov functions provide one of the most direct bridges between classical stability theory and modern reinforcement learning. Broadly, existing approaches use Lyapunov structure in three ways: (i) as a \emph{certificate} for stability/safety (e.g., ROA/level-set reasoning), (ii) as a \emph{constraint} that shapes or restricts policy updates (e.g., CMDP feasibility, projection, shielding), and (iii) as a \emph{learning signal} (e.g., penalties, reward shaping, or empirical risk minimization). These perspectives yield complementary strengths, but they also expose recurring limitations that motivate several key research directions.

\paragraph{Stability certificates: constructing and maintaining valid Lyapunov functions}
A central theme is to interpret a Lyapunov candidate as a stability certificate that induces attractive and invariant level sets. Early work uses this certificate to ensure safe switching among stabilizing control laws while still permitting RL to optimize high-level decisions \cite{perkins2002lyapunov}. A more common modern pathway is to identify the value function (or $Q$-function) with a Lyapunov candidate, leveraging approximate dynamic programming and actor--critic updates with Lyapunov analysis \cite{vamvoudakis2015asymptotically,kamalapurkar2016model,kumar2017diagonal,yao2025lyapunov}. While this creates a principled connection to optimal control (e.g., HJB structure), the resulting guarantees often rely on restrictive assumptions about approximation error, excitation conditions, or the dynamics class (frequently deterministic or structured nonlinear systems). In practice, the learned critic may satisfy Lyapunov-like behavior only on the data distribution encountered during training, raising concerns about certificate \emph{validity under distribution shift} and \emph{robustness to function approximation error}.

\paragraph{Safe-set expansion with learned models and uncertainty: conservatism vs scalability}
A second line of work explicitly learns models and uses Lyapunov level sets to define a safe region of attraction. The GP-based formulation in \cite{berkenkamp2017safe} exemplifies this idea: uncertainty-aware upper bounds are used to enforce a Lyapunov decrease condition over a discretized state set, and the safe region is expanded as data accrue. Related methods incorporate empirical Lyapunov risk into learning objectives \cite{xia2024estimating}. These approaches provide a tangible notion of ``where the agent is safe'', but they also surface two practical bottlenecks: (i) the frequent need for an \emph{initial safe policy or safe set} (otherwise exploration cannot be certified), and (ii) \emph{conservatism} induced by Lipschitz bounds, discretization, or uncertainty envelopes. Moreover, many uncertainty quantification mechanisms (e.g., GP kernels) can be hard to scale and tune in high-dimensional state spaces, which remains a barrier to deployment in complex robotic domains.

\paragraph{CMDP and constrained optimization: feasibility, projections, and bias}
Compared to control-affine stability analyses, Lyapunov-based CMDP formulations provide a unifying RL interface for safety constraints. The Lyapunov-constrained CMDP viewpoint in \cite{chow2018lyapunov} introduces a Bellman-style feasibility condition and yields safe dynamic programming and RL algorithms; extensions include continuous-action policy optimization via parameter projection ($\theta$-projection) and action projection ($a$-projection) \cite{chow2019lyapunov}. Closely related ideas use Lyapunov drift constraints in a QP-like projection of the RL action \cite{cao2023toward}, and shielding mechanisms construct safe policies and exploration strategies using Lyapunov-induced safe sets \cite{huh2020safe}. While these methods can enforce \emph{hard} safety at execution time, they also introduce new failure modes: the feasible set may be empty or numerically fragile, projections may bias gradients and slow convergence, and aggressive constraints can suppress exploration, often yielding safe but overly conservative policies.

\begin{sidewaystable}[htpb]
	\centering
	\small
	\caption{Classification of Lyapunov-function-based safe RL.}
	\label{tab:lyap_rl_classification}
	\setlength{\tabcolsep}{4pt}
	\renewcommand{\arraystretch}{1.05}
	\begin{tabularx}{\linewidth}{@{}p{0.22\linewidth}p{0.18\linewidth}X X X@{}}
			\toprule
			\textbf{Class} & \textbf{Methodology} & \textbf{Works} & \textbf{Pros} &\textbf{Cons} \\
			\midrule
			Switching CLF controllers & RL switches among stabilizing modes & \cite{perkins2002lyapunov} &Safety inherits from each mode; faster learning. &Needs library of stabilizers; limited expressivity. \\
			
			Value-as-Lyapunov (ADP/HJB) & Use $V$ or $Q$ as Lyapunov candidate & \cite{vamvoudakis2015asymptotically,kamalapurkar2016model,kumar2017diagonal,yao2025lyapunov} &Tight link to optimal control; stability proofs under assumptions. &Strong assumptions; approximation error sensitivity; constraints need extra handling. \\
			
			Reachability CLVF & CLVF/safe set via HJR reach--avoid & \cite{lopez2024decomposing} &Strong constraint semantics; interpretable sets. &Poor scaling with dimension; conservative sets. \\
			
			Model+ROA expansion (uncertainty) & Learn model (e.g., GP), enforce Lyapunov decrease over ROA & \cite{berkenkamp2017safe,xia2024estimating} &Explicit safe region; can expand with data. &Often needs safe initializer; Lipschitz/discretization conservatism; GP scaling. \\
			
			CMDP Lyapunov feasibility & Lyapunov constructs feasible set in CMDP & \cite{chow2018lyapunov,chow2019lyapunov} &RL-native constrained framework; DP/PG variants. &Feasibility brittleness; conservative; projection bias. \\
			
			Projection / shielding layers & Project action/params to satisfy Lyapunov constraints & \cite{chow2019lyapunov,cao2023toward,huh2020safe} &Harder safety at execution; modular. &QP/projection cost; feasibility; reduced exploration. \\
			
			Reward shaping (Lyapunov) & Add Lyapunov temporal--difference shaping & \cite{ng1999policy,dong2020principled,ugurlu2025lyapunov} &Simple; can accelerate learning. &Usually soft safety; weight tuning; can distort optimality. \\
			
			Self--learned Lyapunov critics & Learn $V_\theta$ via empirical Lyapunov risk & \cite{liu2020almost,chang2021stabilizing,mccutcheon2025neural,hao2024lyapunov} &Model--light; adaptive certificates. &Drift estimation hard; noisy finite-differences; OOD generalization risk. \\
			
			Physics--informed / Zubov critic & Critic structured as maximal Lyapunov/Zubov & \cite{wang2024actor,zubov1964methods} &Strong inductive bias; robust ROA reasoning. &Coupled/stiff training; approximation reliance. \\
			
			Two--stage CLF + violation loss & Learn CLF then penalize violations in RL & \cite{10155918} &Simple pipeline; reduces unsafe rollouts after CLF. &Soft safety unless filtered; compounding errors. \\
			
			Multi--agent Lyapunov drift & Drift penalties in MARL objectives & \cite{kumar2023task} &Extends to distributed settings. &Non--stationarity; coupled constraints; weaker guarantees. \\
			\bottomrule
		\end{tabularx}
\end{sidewaystable}

\paragraph{Learning signals: penalties, shaping, and self--learned Lyapunov critics}
Several works relax hard constraints and instead use Lyapunov structure as a learning signal. Reward shaping methods add Lyapunov--inspired temporal difference terms to accelerate learning \cite{dong2020principled,ugurlu2025lyapunov,ng1999policy}. These techniques are attractive due to their simplicity and compatibility with standard deep RL pipelines, but safety becomes ``soft'': violations may still occur, and performance is sensitive to shaping weights and discounting assumptions. A more ambitious direction is to learn the Lyapunov function itself from data by minimizing empirical Lyapunov risk under relaxed (``almost'') Lyapunov conditions \cite{liu2020almost,chang2021stabilizing,mccutcheon2025neural,hao2024lyapunov}. In this family, the key technical difficulty is reliably estimating Lyapunov drift/Lie derivatives in a model--free setting, where finite--difference approximations can be noisy and biased. As a result, stability assurances may be empirical or probabilistic, and learned certificates can generalize poorly beyond the training distribution.

\paragraph{Physics--informed and maximal Lyapunov structures}
Recent work explores injecting stronger inductive bias into the critic by adopting physics--informed structures such as Zubov/maximal Lyapunov functions \cite{wang2024actor,zubov1964methods}. These approaches aim to approximate ``largest'' regions of attraction and iteratively update actor and critic to satisfy Lyapunov conditions. Although promising for robustness and interpretability, such methods can lead to stiff coupled optimization (actor improvement must not destroy certificate validity), and success can depend strongly on how Lyapunov constraints are enforced during learning.

\paragraph{Extensions: inaccurate models, two--stage pipelines, and multi--agent settings}
Beyond standard single--agent settings, CLF dissipation constraints with imperfect models provide a pragmatic compromise: stability is encouraged/enforced through minimum dissipation constraints while using model--free policy optimization \cite{westenbroek2020learning,westenbroek2022lyapunov}. Two-stage pipelines that first learn a CLF (and possibly dynamics) and then penalize Lyapunov violations during RL training have also been reported \cite{10155918}. Finally, Lyapunov drift has been incorporated into multi-agent actor--critic objectives to stabilize distributed learning and coordination \cite{kumar2023task}, though non-stationarity and credit assignment complicate both certificate learning and theoretical analysis.

%%%%%%%%%%%%%%%%%%%%%%%%%%%%%%%%%%%%%%%%%%%%%%%%%%%%%%%%%%%%%%%%%%%%%%%%%%%%%%%%%%%%%%%%%%%%%%%%%%%%%%%%%%%%%%%%%%%

\subsection{Key Research Issues and Open Problems}
\label{subsec:lyap_open_directions}

\paragraph{Reducing dependence on an initial safe policy or safe set}
Many provably safe schemes assume a known safe baseline (or a seed ROA) to bootstrap learning \cite{berkenkamp2017safe,chow2019lyapunov}. Developing \emph{safe exploration without a certified initializer} remains a critical gap, particularly for real robots where resets and failures are costly.

\paragraph{Certificate validity under function approximation and distribution shift}
Deep function approximators can produce Lyapunov candidates that appear stable on-policy but fail off-policy. Robust training objectives, explicit generalization tests for certificates, and distributionally robust formulations for Lyapunov decrease constraints are needed, especially in high-dimensional continuous control \cite{chang2021stabilizing,westenbroek2020learning}.

\paragraph{Reliable estimation of Lyapunov drift/Lie derivatives in model-free settings}
Finite-difference approximations used in self-learned Lyapunov critics can be sensitive to noise and step size \cite{chang2021stabilizing}. Open problems include variance reduction, uncertainty-aware drift estimation, and hybrid approaches that exploit partial models or learned latent dynamics while maintaining valid bounds \cite{berkenkamp2017safe}.

\paragraph{Conservatism, feasibility, and the safety performance trade-off}
Hard constraints via projection/shielding can become overly conservative or even infeasible \cite{chow2019lyapunov,cao2023toward,huh2020safe}. A major practical challenge is designing \emph{adaptive} constraint tightening/relaxation strategies and diagnosing when conservatism stems from certificate choice, discretization/Lipschitz bounds, or approximation error.

\paragraph{Scaling certificates and uncertainty quantification}
Reachability-based constructions yield strong semantics but scale poorly with dimension \cite{lopez2024decomposing}. GP-based uncertainty methods can be difficult to scale and tune \cite{berkenkamp2017safe}. Promising directions include compositional certificates, structured Lyapunov parameterizations, and scalable uncertainty surrogates that preserve conservative bounds.

\paragraph{Unifying CLF/QP-style filters with deep RL training dynamics}
Action filters and QPs can guarantee safety during execution, but their interaction with policy gradients can introduce bias and learning instability \cite{chow2019lyapunov,cao2023toward}. Better theory and practice are needed for end-to-end training with differentiable safety layers and for quantifying how projection affects optimality and convergence.

\paragraph{Multi-agent and partially observed environments}
Lyapunov drift augmentation in MARL is promising \cite{kumar2023task}, but certificates become harder to define when agents induce non-stationary dynamics or when safety is coupled across agents. Incorporating memory and uncertainty models (e.g., Transformer-based risk prediction) in CMDPs \cite{jeddi2021lyapunov} suggests a path forward, but scalable theory for partial observability and joint safety constraints is still underdeveloped.

Table~\ref{tab:CLF} summarizes the literature reviewed for use of Lyapunov functions in RL. It can be observed that four formal approaches to incorporate Lyapunov functions with RL involve: (i) Lyapunov-based reward-shaping; (ii) CLF-RL-QP formulations; (iii) candidate Lyapunov function as the value/critic function; (iv) Lyapunov/CLF-based loss function. 

%%%%%%%%%%%%%%%%%%%%%%%%%%%%%%%%%%%%%%%%%%%%%%%%%%%%%%%%%%%%%%%%%%%%%%%%%%%%%%%%%%%%%%%%%%%%%%%%%%%%%%%%%%%%%%%%%%%
%%%%%%%%%%%%%%%%%%%%%%%%%%%%%%%%%%%%%%%%%%%%%%%%%%%%%%%%%%%%%%%%%%%%%%%%%%%%%%%%%%%%%%%%%%%%%%%%%%%%%%%%%%%%%%%%%%%

\section{Barrier Functions for Reinforcement Learning}\label{section:ControlBF}

\begin{sidewaystable}[htpb]
\centering
\caption{Summary of safe RL approaches using barrier functions.}
\label{tab:CBF}
\small
\setlength{\tabcolsep}{4pt}
\renewcommand{\arraystretch}{0.95}
\begin{tabular}{@{}p{0.20\linewidth} p{0.40\linewidth} p{0.18\linewidth} p{0.20\linewidth}@{}}
\toprule
\textbf{Reference} & \textbf{Method} & \textbf{RL algorithm} & \textbf{Simulation} \\
\midrule

Ohnishi et al.\ (2019) \cite{ohnishi2019barrier} & Barrier-certified adaptive RL via GP-SARSA & GP-SARSA & Quadrotor, brushbot \\
Cheng et al.\ (2019) \cite{cheng2019end} & CBF compensation via QP shielding & TRPO, DDPG & Inverted pendulum, car following \\
Marvi \& Kiumarsi (2021) \cite{marvi2021safe} & CBF term in cost-to-go function & ADP & Lane-keeping \\
Emam et al.\ (2021) \cite{emam2021safe} & Robust CBF-QP with GP disturbance estimation & SAC & Robotarium \\
Ma et al.\ (2021) \cite{ma2021model} & Generalized CBF in actor-critic with Lagrangian & Actor-critic & SUMO \\
Cai et al.\ (2021) \cite{cai2021safe} & Decentralized CBF shields for multi-agent & Multi-agent DDPG & Particle-env \\
Zhao et al.\ (2022) \cite{zhao2022safe} & Barrier certificate constraints in policy optimization & DDPG & OpenAI Gym \\
Zhang et al.\ (2022) \cite{zhang2022barrier} & Barrier terms in actor and critic loss & Actor-critic & Formation control (mobile robots) \\
Wang (2022) \cite{wang2022ensuring} & Soft barrier (slack) in CBF-QP filter & PPO & CommonRoad-RL \\
Huang et al.\ (2022) \cite{huang2022barrier} & SOS programming for CBF-RL & DDPG & OpenAI Gym \\
Song et al.\ (2022) \cite{song2022safe} & CBF-QP with online CBF learning & SAC & ROS Gazebo \\
Cheng et al.\ (2023) \cite{cheng2023safe} & Disturbance-observer-based robust CBF-QP & SAC & Unicycle, quadrotor \\
Wang et al.\ (2023) \cite{wang2023enforcing} & Generative-model soft barriers in bilevel optimization & PPO & Cartpole, rocket landing \\
Liu et al.\ (2023) \cite{liu2023safe1} & Discrete exponential CBF constraints & Value iteration & Multi-player game \\
Liu et al.\ (2023) \cite{liu2023safe2} & Discrete barrier reward + constraints & Policy iteration, AC & Inverted pendulum \\
Yang et al.\ (2023) \cite{yang2023model} & Joint NN barrier certificate + safe policy & PPO & Safety Gym, MetaDrive \\
Zhao et al.\ (2023) \cite{zhao2023barrier} & Barrier certificate in policy loss; formal verification & Actor-critic & Power systems \\
Ranjan et al.\ (2024) \cite{ranjan2024barrier} & CBF-based reward shaping & TD3 & OpenAI Gym, Unitree Go1 \\
Sabouni et al.\ (2024) \cite{sabouni2024reinforcement} & Bilevel MPC-CBF + RL parameter learning & Actor-critic, MPC & Multi-agent CAV \\
Hou et al.\ (2024) \cite{hou2024safe} & Disturbance-observer CBF-QP & SAC & Gymnasium \\
Yang et al.\ (2024) \cite{yang2024enhancing} & CBF from human demonstration; shielding in inverse RL & Inverse RL & Demolition Derby, Panda Arm (real) \\
Zhang et al.\ (2024) \cite{zhang2024constrained} & Smooth log barrier in actor loss & SAC & Safety Gym, Unitree A1 \\
Dey et al.\ (2024) \cite{dey2024p2bpo} & Controllable soft barrier penalty & PPO & Safety Gym, Safe MuJoCo \\
Dinh et al.\ (2024) \cite{dinh2024towards} & CBF policy filter with local-search CBF & PPO, DDPG & OpenAI Gym \\
Kalaria et al.\ (2024) \cite{kalaria2024disturbance} & Residual-model + disturbance-observer CBF-QP & PPO & Safety Gym, F1/10 racing \\
Liu et al.\ (2025) \cite{liu2025safe} & Robust neural CBF reward shaping & MADDPG & Robot navigation (custom) \\
Xie et al.\ (2025) \cite{xie2025certificated} & CBF-based reward shaping & Actor-critic & Gymnasium, HoloOcean \\

\bottomrule
\end{tabular}
\end{sidewaystable}

There has been significant recent research on the use of barrier functions with RL to define safe sets for the control policy. Fig.~\ref{fig:classification_barrier} shows a broad classification for the use of barrier function-based approaches in safe RL. Similar to reward-shaping using CLF/Lyapunov terms in Section \ref{section:ControlLF}, reward-shaping using CBF terms is observed in several works to ensure state constraints are satisfied. Marvi et al.~\cite{marvi2021safe} propose adding a CBF term to the cost-to-go function for a safe optimal control problem with nonlinear dynamics. A coefficient $\eta$ (damping term) is introduced along with the CBF term to control the effect of barrier term, this can be interpreted as a trade-off between safety and optimality. The authors also prove the additional barrier term maintains safety and optimality (under the assumption of finite value barrier function and suitable value of $\eta$). The MPC formulation for a discrete time control system is augmented as follows:
\begin{subequations}
	\begin{gather}
		J^{*}(x) = \min_u \biggl\{\tilde{c}_0(x_0,u_0) + \sum_{k=1}^{N-1} \tilde{c}(x_k,u_k) + \tilde{c}_f(x_N) \biggr\} \\ 
		\tilde{c}(x_k,u_k) = c(x_k,u_k) + \eta B_u(u) + \eta B_x(x) \\
		\tilde{c}_0(x_0,u_0) = c_0(x_0,u_0) + \eta B_u(u) \\
		\tilde{c}_f(x_N) = c_f(x_N) + \eta B_f(x) 
	\end{gather}
\end{subequations}
where, $c_0(\cdot)$, $c(\cdot)$ and $c_f(\cdot)$ are the initial, stage and terminal cost, respectively. $\eta \geq 0$ is the barrier damping coefficient. $B_x(\cdot)$, $B_u(\cdot)$ and $B_f(\cdot)$ are barrier functions for constraints on states, control input and terminal conditions. An actor-critic algorithm is used to prove efficacy of the proposed approach for a lane-keeping problem. Authors in \cite{liu2023safe2} propose a similar reward-shaping where the reward function is augmented with an exponential CBF term. A policy iteration (PI) and actor-critic RL framework is considered for an inverted pendulum environment. Reward-shaping is further explored in \cite{ranjan2024barrier, liu2025safe, xie2025certificated} considering exponential, quadratic and robust neural CBFs. 

\emph{Cart-Pole instantiation.} On Cart-Pole (Section~\ref{subsec:cartpole_example}), the CBF--QP filter computes $u^{\text{safe}} = \arg\min_u \|u - u^{\text{RL}}\|^2$ subject to $L_f h_i(x) + L_g h_i(x) u + \alpha(h_i(x)) \geq 0$ for $i \in \{1,2\}$, projecting the RL action onto the intersection of position-safe and angle-safe sets at every time step.

In~\cite{ohnishi2019barrier}, authors propose a barrier-certified adaptive RL for a quadrotor control and brushbot navigation application using Gaussian process (GP) SARSA algorithm. The proposed approach follows a few key steps: (i) Adaptive model learning (ii) Structured model learning using sparse optimization satisfying monotone approximation property (Section III-B of cited work) in a defined reproducing kernel Hilbert space (RKHS), to capture the structure of agent dynamics (useful for computing Lie derivative of control barrier functions); (iii) Adaptive action-value function approximation with barrier-certified policy updates (Section IV-A). The authors prove global optimality of solutions to update the control policy. Closely related to GP/RKHS-based barrier learning, Brunke and Zhou~\cite{brunke2022barrier} introduce barrier Bayesian linear regression (BBLR), where a Bayesian linear model with engineered features online-learns the unknown drift/input dynamics that enter the CBF condition. Posterior credible bounds tighten the CBF inequalities used in the QP, yielding probabilistic safety guarantees while avoiding the cubic per-step cost of full GP regression. BBLR is a practical compromise between full GP-SARSA~\cite{ohnishi2019barrier} and observer-based disturbance compensation~\cite{cheng2023safe,hou2024safe}. Cheng et al.~\cite{cheng2019end} propose a policy combining model-free RL and model-based CBF. The algorithm is tested on inverted pendulum and car-following with vehicle-to-vehicle communication using TRPO and DDPG based RL. The authors propose two approaches combining CBF and RL controller: (i) The first uses a concept similar to CBF shielding (only compensates for RL controller in case of constraint violation and formulated by solving a QP at each iteration) given as,
\begin{equation}
	u_k(x) = u_k^{RL}(x) + u_k^{CBF}(x, u_k^{RL}) \label{eqn:CBF_QP_u}
\end{equation}
where, $u_k^{RL}(\cdot)$ is the computed RL policy and $u_k^{CBF}(\cdot)$ is computed using the minimum control magnitude QP formulation (\ref{eqn:CBF_QP}); (ii) The second approach uses prior CBF controllers formulated through solving a QP along with the CBF shielding and RL controller given as,
\begin{equation}
	u_k(x) = u_k^{RL}(x) + \sum_{j=0}^{k-1}u_j^{CBF}(x, u_0^{RL}, \dots, u_{k-1}^{RL} ) \nonumber\\
	+ u_k^{CBF}(x, u_k^{RL} + \sum_{j=0}^{k-1}u_j^{CBF} ) 
\end{equation}
In this case, dependence on all prior CBF controller enhances learning efficiency. The second approach guides exploration as well by taking prior computed CBFs into consideration. Probabilistic guarantees are provided for TRPO and DDPG algorithms implemented. In~\cite{ma2021model}, authors propose a generalized CBF for a model-based constrained RL policy optimization algorithm. The problem is formulated for an actor-critic algorithm with the actor policy update as an optimization problem with CBF constraints. Lagrange multipliers are used to reformulate the problem as an unconstrained optimization problem along with a distance constraint (collision avoidance simulations). Furthermore, a parameter $\xi$ (constraint violation metric) is computed to compare with a predetermined threshold for updating the parameters of CBF to adjust constraint satisfaction. A robust CBF is proposed in~\cite{emam2021safe} for a nonlinear system with disturbance. The disturbance is approximated using a GP (this approximated model is used to compute the Lie derivatives in (\ref{eqn:CBF_QP}) and the robust CBF is combined with RL actions to form a relaxed QP problem). The slack variable is introduced to avoid deadlock when no safe action is available. The proposed approach is tested on unicycle and car following environments using a SAC-Robust CBF formulation. Similar optimization formulations are observed in \cite{huang2022barrier,song2022safe,cheng2023safe, yang2024enhancing, dinh2024towards, kalaria2024disturbance}. Soft-barrier constraints are proposed in \cite{wang2022ensuring}, where a slack variable $\delta$ is introduced to relax the barrier constraints and ensure a feasible solution for the CBF-RL-QP formulation. 
A multi-agent RL with decentralized CBF shields is proposed by authors in~\cite{cai2021safe} for a collision-avoidance problem. The two types of decentralized CBF shields are formulated namely, cooperative and non-cooperative, the safe policy is computed as a constrained optimization problem for the two types of CBF shields and action by the RL agent, similar to (\ref{eqn:CBF_QP_u}). A multi-agent DDPG-CBF formulation is tested on a two patrolman task. 
In~\cite{cheng2023safe}, authors use disturbance observers and CBFs with RL to provide safety guarantees. A policy is sampled from the RL agent followed by disturbance estimation from the disturbance-based observer, similar to prior approaches discussed Lie derivative is approximated using disturbance-based observers. A QP (\ref{eqn:CBF_QP_u}) is solved using the CBF as constraints and the action sampled by the RL agent to solve for the safe action. SAC algorithm is used for simulations on a 2D quadrotor and a unicycle model. Hou et al. \cite{hou2024safe} and Kalaria et al. \cite{kalaria2024disturbance} follow a similar disturbance-observer-based CBF filter to enforce hard constraints. 

\begin{table*}[htpb]
	\centering
	\small
	\caption{Classification of barrier-function-based safe RL.}
	\label{tab:cbf_rl_classification_compact}
	\setlength{\tabcolsep}{4pt}
	\renewcommand{\arraystretch}{1.05}
	\begin{tabularx}{\linewidth}{@{}p{0.22\linewidth}p{0.18\linewidth}X X X@{}}
			\toprule
			\textbf{Class} & \textbf{Methodology} & \textbf{Works} & \textbf{Pros} &\textbf{Cons}\\
			\midrule
			CBF reward shaping & Barrier terms in reward/cost & \cite{marvi2021safe,liu2023safe2,ranjan2024barrier,liu2025safe,xie2025certificated} &Simple integration; helps exploration. &Soft safety; sensitive weights; may be over-conservative. \\
			
			Barrier-certified learning (GP/RKHS) & Learn structure to compute Lie derivatives; certify updates & \cite{ohnishi2019barrier} &Uncertainty-aware certificates; strong theory (assumption-heavy).&GP/RKHS scaling; conservatism; derivative estimation burden. \\
			
			CBF-QP shielding (filter) & $u^{safe}=\arg\min\|u-u^{RL}\|$ s.t.\ CBF & \cite{cheng2019end,emam2021safe,huang2022barrier,song2022safe,cheng2023safe,yang2024enhancing,dinh2024towards,hou2024safe,kalaria2024disturbance} &Hard constraint enforcement when feasible; modular. &Needs model/derivatives; QP cost; feasibility/deadlock; exploration restriction. \\
			
			Soft CBF / slack QP & Slack variables to guarantee feasibility & \cite{wang2022ensuring,emam2021safe} &Avoids infeasibility/deadlock. &Safety becomes graded; slack tuning; may hide modeling error. \\
			
			Constrained policy optimization & Actor update constrained by CBF; Lagrange/primal-dual & \cite{ma2021model,zhang2022barrier,liu2023safe1,yang2023model,dey2024p2bpo,zhang2024constrained} &End-to-end safety-aware learning; can reduce per-step QP. &Often expectation/approx.\ safety; multiplier tuning; non-convexity. \\
			
			Learner--verifier frameworks & Barrier loss + formal verification step & \cite{zhao2022safe} &Stronger assurance via verification. &Verification overhead; scalability depends on verifier. \\
			
			Robust/disturbance-aware CBFs & GP disturbance models/observers for robust CBF-QP & \cite{emam2021safe,cheng2023safe,hou2024safe,kalaria2024disturbance} &Better under perturbations/unmodeled effects. &Estimator error weakens guarantees; added tuning/latency. \\
			
			Multi-agent decentralized shields & Local CBF shields (coop/non-coop) & \cite{cai2021safe} &Distributed collision avoidance; modular. &Coupled feasibility; non-stationarity; network assumptions. \\
			
			Chance constraints / bilevel & Probabilistic safety via bilevel soft barriers & \cite{wang2023enforcing,sabouni2024reinforcement} &Targets stochastic safety directly. &High compute; calibration sensitivity; bilevel complexity. \\
			\bottomrule
		\end{tabularx}
\end{table*}

Similar to CLF/Lyapunov violations considered in policy loss function, CBF terms are augmented to the loss functions in barrier-based safe RL approaches. Zhao et al. \cite{zhao2022safe} propose a learner-verifier framework where barrier terms are added to the actor loss function and further actions taken by the agent are formally verified by solving an optimization for each barrier certificate. Authors in \cite{zhang2022barrier, liu2023safe1,yang2023model, dey2024p2bpo, zhang2024constrained} follow the Lagrangian method to incorporate the barrier constraints in the actor objective function. 

Authors in~\cite{wang2023enforcing} use soft barrier constraints through generative-model-based RL in an unknown stochastic environment. Their approach forces hard safety chance constraints (probabilistic safety constraints) using bilevel optimization formulation. The lower problem encodes hard safety constraints with a generative-model-based soft barrier function and the upper problem maximizes the total expected return of the policy. One limitation of this approach is the high computational complexity of the generative model, as stated by the authors. Sabouni et al. \cite{sabouni2024reinforcement} follow a bilevel optimization architecture as well. However, RL is used to learn the optimal parameters of a MPC formulation.  

It can be seen that most approaches concerning use of barrier functions follow a relaxed QP formulation to act as a compensating controller. The key differences in approaches are how the nonlinear dynamics are approximated to compute the safe set (forward invariant set). System information becomes necessary to compute these safe sets and conservative policies due to restricted exploration are observed. Table~\ref{tab:CBF} summarizes the literature reviewed for use of barrier functions in safe RL.
%%%%%%%%%%%%%%%%%%%%%%%%%%%%%%%%%%%%%%%%%%%%%%%%%%%%%%%%%%%%%%%%%%%%%%%%%%%%%%%%%%%%%%%%%%%%%%%%%%%%%%%%%%%%%%%%%%%

\subsection{Classification and Discussion}
\label{subsec:cbf_discussion_open_problems}

Table \ref{tab:cbf_rl_classification_compact} provides a tabular comparison of each methodology and its shortcomings. Barrier-function-based safe RL methods have grown rapidly due to their ability to formalize safety as \emph{forward invariance} of a set, i.e., trajectories remain in a constraint-admissible region when a control barrier function (CBF) condition is enforced. The surveyed literature can be organized around how the barrier enters the learning loop: (i) as a \emph{soft learning signal} (reward/cost shaping), (ii) as a \emph{hard execution-time filter} (CBF-QP shielding), and (iii) as a \emph{constraint embedded into policy optimization} (Lagrangian/learner--verifier/bilevel formulations). These categories provide complementary trade-offs between scalability, theoretical rigor, and practical deployability.

\paragraph{Barrier shaping: simplicity with limited guarantee strength}
Reward shaping with barrier terms modifies the cost-to-go or reward to discourage constraint violation while preserving performance objectives \cite{marvi2021safe,liu2023safe2,ranjan2024barrier,liu2025safe,xie2025certificated}. This approach is attractive because it integrates seamlessly with standard actor--critic or policy iteration pipelines and can improve learning efficiency by biasing exploration away from unsafe states. However, shaping is typically a \emph{soft} mechanism: unless the shaped objective is paired with additional constraint-enforcement machinery, safety violations can still occur particularly during early exploration or under distribution shift. Moreover, barrier weights introduce a brittle safety--optimality trade-off; poorly tuned penalties can either fail to prevent violations or over-regularize and yield conservative policies.

\paragraph{CBF shielding via QP filters: hard constraints with model dependence and feasibility challenges}
A dominant template in barrier-based safe RL is to treat the RL policy as a nominal controller $u^{RL}$ and compute a minimally modified safe action via a CBF-QP filter \cite{cheng2019end,emam2021safe,huang2022barrier,song2022safe,cheng2023safe,yang2024enhancing,dinh2024towards,kalaria2024disturbance,hou2024safe}. This ``safety layer'' can provide strong practical safety because it enforces barrier constraints at execution time whenever the QP is feasible. The main limitation is that computing the CBF constraints requires system information typically a control-affine structure and Lie derivative estimates, so most methods introduce a model learner (e.g., GP) or an estimator/observer to approximate unknown dynamics \cite{emam2021safe,cheng2023safe,kalaria2024disturbance}. This creates a tight coupling between certificate validity and modeling accuracy. In addition, QP feasibility (and deadlock near boundaries) remains a core issue. Slack-variable relaxations address feasibility \cite{wang2022ensuring}, but transform strict safety into a penalty violation term that must be tuned and monitored.

\paragraph{Constrained policy optimization and verification: end-to-end training vs  enforceability}
Another class of approaches incorporates barrier constraints directly into the policy update, often through Lagrangian formulations or learner--verifier pipelines \cite{zhao2022safe,zhang2022barrier,liu2023safe1,yang2023model,dey2024p2bpo,zhang2024constrained}. These methods can reduce reliance on per-step QPs and enable differentiable end-to-end training, but they frequently enforce safety \emph{in expectation} or via approximate penalty mechanisms. Formal verification steps can strengthen guarantees \cite{zhao2022safe}, yet they introduce computational overhead and typically depend on the tractability of verifying barrier certificates. In stochastic environments, bilevel and chance-constrained formulations seek to capture probabilistic safety more directly \cite{wang2023enforcing,sabouni2024reinforcement}, but at the cost of substantial computational complexity and sensitivity to probabilistic model calibration.

\paragraph{Multi-agent and disturbance-aware settings: realistic but significantly harder}
Realistic deployment often entails disturbances, unmodeled dynamics, and multi-agent coupling. Disturbance-aware CBF filters that leverage GPs or disturbance observers aim to make invariance robust to perturbations \cite{emam2021safe,cheng2023safe,hou2024safe,kalaria2024disturbance}. Meanwhile, decentralized CBF shielding extends the QP-filter template to multi-agent collision avoidance \cite{cai2021safe}. These extensions improve realism but amplify fundamental difficulties: estimating Lie derivatives reliably under uncertainty, maintaining feasibility under coupled constraints, and ensuring stability of the combined estimator--filter--RL closed loop.
%%%%%%%%%%%%%%%%%%%%%%%%%%%%%%%%%%%%%%%%%%%%%%%%%%%%%%%%%%%%%%%%%%%%%%%%%%%%%%%%%%%%%%%%%%%%%%%%%%%%%%%%%%%%%%%%%%%
\subsection{Key Issues and Open Problems}
\label{subsec:cbf_open_directions}

\paragraph{Model dependence and certificate validity under approximation}
Most CBF guarantees hinge on accurate Lie derivative computation. When dynamics are unknown, learned models/observers introduce approximation error that can invalidate safety certificates. A core research need is \emph{conservative-yet-nontrivial uncertainty bounds} for learned derivatives that scale to high dimensions, enabling certificates that remain valid under model error.

\paragraph{Feasibility, deadlock, and principled slack allocation}
QP filters can become infeasible near constraint boundaries or under conflicting constraints, causing deadlock. Slack variables improve feasibility \cite{wang2022ensuring}, but selecting slack penalties is ad hoc and may conceal systematic model error. Developing principled \emph{violation penalty terms} (state-dependent slack penalties, risk-aware slack allocation, and safety margin adaptation) is a key direction.

\paragraph{Exploration under hard safety constraints}
CBF shielding prevents unsafe actions but can severely restrict exploration, especially when the safe set is small or conservative due to modeling bounds. Methods that \emph{actively expand} the safe set while exploring (e.g., safe set enlargement strategies, optimistic exploration within certified regions, or barrier-aware intrinsic motivation) remain underdeveloped compared to Lyapunov ROA expansion.

\paragraph{End-to-end differentiable safety layers and learning stability}
Integrating QP filters into backpropagation (differentiable QPs) can reduce projection-induced bias and improve sample efficiency, but raises stability and computational challenges. Understanding how safety layers affect policy gradients, convergence, and optimality gaps is an open problem.

\paragraph{Chance constraints, distribution shift, and calibration}
Probabilistic safety is essential in stochastic environments, but chance-constrained/bilevel approaches depend on calibrated uncertainty models \cite{wang2023enforcing}. Research is needed on robust chance constraints under miscalibration, offline-to-online transfer, and rare-event safety validation.

\paragraph{Multi-agent coupled constraints and partial observability}
Decentralized CBF shields \cite{cai2021safe} are promising, yet coupled constraints yield complex feasibility regions and non-stationary learning dynamics. Open directions include compositional multi-agent barriers, scalable coordination protocols, and barrier certificates under partial observability and intermittent communication.

Barrier-function-based safe RL offers a pragmatic path to enforcing state constraints through forward invariance, with CBF-QP shielding emerging as a dominant tool. Table~\ref{tab:CBF} summarizes the literature reviewed for the use of barrier functions in RL. The central trade-off mirrors Lyapunov-based methods: strong safety enforcement often requires substantial system knowledge and can induce conservatism, while purely learning-based barrier objectives scale more easily but provide weaker guarantees. Bridging this gap via scalable uncertainty quantification, feasibility-aware filters, and exploration that expands certified safe sets appears to be the key to widely deployable barrier-based safe RL.

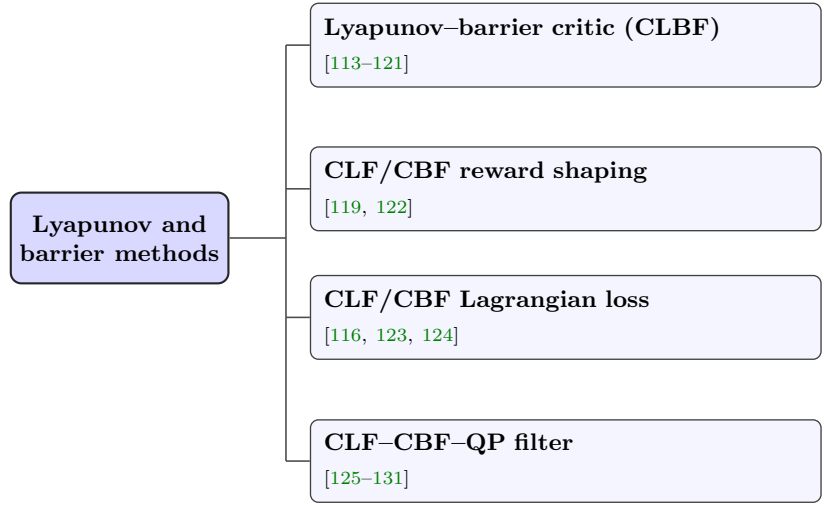
\begin{figure}[t]
\centering
\begin{tikzpicture}[
  >=stealth, line width=0.5pt,
  rootnode/.style={
    rectangle, rounded corners=4pt, draw=black!85, line width=0.8pt,
    fill=blue!15, minimum width=2.8cm, minimum height=1.2cm,
    align=center, font=\bfseries\small
  },
  catnode/.style={
    rectangle, rounded corners=3pt, draw=black!75, line width=0.5pt,
    fill=blue!4, text width=6.4cm, align=left, inner sep=5pt, font=\small
  },
  branch/.style={line width=0.6pt, draw=black!70}
]
  \node[rootnode] (root) at (0, 0) {Lyapunov and\\barrier methods};

  \node[catnode] (c1) at (5.9, 2.55) {%
    \textbf{Lyapunov--barrier critic (CLBF)}\\[2pt]%
    \footnotesize \cite{9827972,du2023reinforcement,cohen2023safe,zhao2024nlbac,mandal2024formally,wei2024asymmetric,wang2024control,zhang2023robust,tan2025hierarchical}};

  \node[catnode] (c2) at (5.9, 0.65) {%
    \textbf{CLF/CBF reward shaping}\\[2pt]%
    \footnotesize \cite{mizuta2024cobl,wang2024control}};

  \node[catnode] (c3) at (5.9, -1.05) {%
    \textbf{CLF/CBF Lagrangian loss}\\[2pt]%
    \footnotesize \cite{zhao2024nlbac,cocaul2024safe,zhao2025nlbac}};

  \node[catnode] (c4) at (5.9, -2.95) {%
    \textbf{CLF--CBF--QP filter}\\[2pt]%
    \footnotesize \cite{li2019temporal,choi2020reinforcement,meng2023reinforcement,jang2024safe,chen2024collision,tan2026adaptive,tan2025fixed}};

  \coordinate (j) at (2.2, 0);
  \draw[branch] (root.east) -- (j);
  \draw[branch] (j |- c1.west) -- (j |- c4.west);
  \draw[branch] (j |- c1.west) -- (c1.west);
  \draw[branch] (j |- c2.west) -- (c2.west);
  \draw[branch] (j |- c3.west) -- (c3.west);
  \draw[branch] (j |- c4.west) -- (c4.west);
\end{tikzpicture}
\caption{Classification of methods combining Lyapunov and barrier functions in safe RL. Categories correspond to how the joint CLF--CBF structure is embedded: as a single Lyapunov--barrier function (CLBF) in the critic, as reward shaping, as a Lagrangian loss term, or as an execution-time QP safety filter.}
\label{fig:classification_barrier_lyapunov}
\end{figure}

%%%%%%%%%%%%%%%%%%%%%%%%%%%%%%%%%%%%%%%%%%%%%%%%%%%%%%%%%%%%%%%%%%%%%%%%%%%%%%%%%%%%%%%%%%%%%%%%%%%%%%%%%%%%%%%%%%%
%%%%%%%%%%%%%%%%%%%%%%%%%%%%%%%%%%%%%%%%%%%%%%%%%%%%%%%%%%%%%%%%%%%%%%%%%%%%%%%%%%%%%%%%%%%%%%%%%%%%%%%%%%%%%%%%%%%

\section{Lyapunov and Barrier Functions for Reinforcement Learning}\label{section:CLFandCBF}
Few recent approaches have leveraged the stability guarantees provided by Lyapunov functions and the constraint satisfaction capability of barrier functions to form a single approach. Fig.~\ref{fig:classification_barrier_lyapunov} shows a broad classification of approaches using Lyapunov and barrier methods. In~\cite{li2019temporal}, the authors propose the use of temporal logic based RL using CBFs for the PPO algorithm in a dynamic obstacle avoidance problem. The authors consider a syntactically co-safe truncated temporal logic which is used to generate a finite state automaton (FSA), this generated FSA is then augmented with the MDP to introduce temporal logic in the RL formulation. A CLF-CBF-QP (Proposition 1 in Section III of cited work) essentially formulates the CLF and CBF as constraints for a QP problem with a quadratic objective to compute the safe action. This formulated QP is relaxed using $\delta$ to avoid deadlock between CLF and CBF constraints. The CLF-CBF-QP for a general continuous time nonlinear control affine dynamical system is given as follows:
\begin{subequations}
	\begin{gather}
		\min_{u \in \mathcal{U}} \Vert u \Vert^2 + k\delta \\
		\text{s.t.  } L_f V(x) + L_g V(x)(u_{RL} + u) \leq \delta -\alpha V(x) \\
		L_f h(x) + L_g h(x)(u_{RL} + u) + \alpha(h(x))\leq 0 \\
		\delta \geq 0
	\end{gather}
\end{subequations}

\emph{Cart-Pole instantiation.} The CLF--CBF--QP on Cart-Pole simultaneously enforces Lyapunov decrease through $V$ (driving toward the upright equilibrium) and barrier non-decrease through $h_1, h_2$ (keeping the cart on the track and the pole upright). The slack $\delta$ becomes nontrivial when these objectives conflict near the track boundary, where stabilization may temporarily require violating Lyapunov decrease to satisfy invariance.

\begin{sidewaystable}[htpb]
\centering
\caption{Summary of safe RL approaches with Lyapunov and Barrier functions}\label{tab:CBF_CLF}
\setlength{\tabcolsep}{4pt}
\renewcommand{\arraystretch}{1.2}
\begin{tabular}{@{}p{0.15\linewidth} p{0.33\linewidth} p{0.24\linewidth} p{0.24\linewidth}@{}}
\toprule
\textbf{Reference} & \textbf{Method} & \textbf{Algorithm} & \textbf{Simulation} \\
\midrule
Li et al.~(2019)~\cite{li2019temporal}       & TL-guided CLF-CBF-QP policy optimization          & PPO             & Unicycle \\
Choi et al.~(2020)~\cite{choi2020reinforcement} & CLF-CBF-QP policy optimization                 & DDPG            & Bipedal robot \\
Zhang et al.~(2022)~\cite{9827972}           & CLF-CBF constraints for policy optimization       & ADP             & Four-wheel vehicle \\
Du et al.~(2023)~\cite{du2023reinforcement}  & Control Lyapunov-Barrier function as value fn.    & Actor-critic    & 2D quadrotor \\
Cohen et al.~(2023)~\cite{cohen2023safe}     & Lyapunov-like CBF                                 & ADP             & Collision avoidance \\
Zhao et al.~(2023)~\cite{zhao2023stable}     & CLF-CBF constraints as Lagrangian loss            & Actor-critic    & Unicycle \\
Meng et al.~(2023)~\cite{meng2023reinforcement} & CLF-CDBF-TCBF constraints in RL-QP            & DDPG            & CarSim \\
Jang et al.~(2024)~\cite{jang2024safe}       & Lyapunov-based CBF in QP formulation              & LQR             & Multirotor, CartPole, rover \\
Wei et al.~(2024)~\cite{wei2024asymmetric}   & Asymmetric time-varying integral BLF backstepping & Actor-critic    & Pendulum, one-link robot \\
Mizuta et al.~(2024)~\cite{mizuta2024cobl}   & CLF-CBF based reward shaping                      & Diffusion model & Custom \\
Zhao et al.~(2024)~\cite{zhao2024nlbac}      & CLF-CBF Lagrangian loss                           & SAC             & Simulated car following \\
Wang et al.~(2024)~\cite{wang2024control}    & Control Lyapunov-Barrier augmented rewards        & Policy iteration & Chemical process \\
Mandal et al.~(2024)~\cite{mandal2024formally} & Formal verification of neural LBC               & PPO             & Spacecraft control \\
Chen et al.~(2024)~\cite{chen2024collision}  & CLF-CBF-QP low-level + DQN high-level controller  & DQN             & Simulink autonomous driving \\
Cocaul et al.~(2024)~\cite{cocaul2024safe}   & Lyapunov risk term and barrier in critic loss     & PPO             & OpenAI Gym \\
Zhao et al.~(2025)~\cite{zhao2025nlbac}      & Lagrangian CLF-CBF augmented policy loss          & Actor-critic    & OpenAI Gym, Safe-Control-Gym \\
\hline
\end{tabular}
\end{sidewaystable}

where, $V(\cdot)$ and $h(\cdot)$ denote the candidate Lyapunov and barrier functions, respectively. $u_{RL}$ denotes the computed action by the RL agent which is then compensated by adding the CLF and CBF constraints using the QP formulation. FSA augmented MDP provides reward for RL, hard constraints for CBF to guide exploration and provide safe set for CLF. The overall control of the RL agent is augmented as follows:
\begin{equation}
	u_k(x) = u_k^{RL} + u_k^{\text{CLF-CBF}}(x,u_k^{RL})
\end{equation}
The approach combines model-based planning (temporal logic) and model-free control (RL). Choi et al.~\cite{choi2020reinforcement} use RL to learn model uncertainty for CLF and CBF dynamic constraints for a nonlinear bipedal robot (assumption: input-output linearization) to address the issue of model uncertainty in data-driven RL. A QP formulation with CLF and CBF constraints (similar to~\cite{li2019temporal}) is used to solve for the control. A DDPG agent is used for learning uncertainty terms in CLF, CBF and other dynamic constraints (policy for RL). Authors in \cite{meng2023reinforcement} follow a similar CLF-CBF-RL-QP formulation by considering CLF, control dependent (CDBF) and time-varying CBF (TCBF). In~\cite{jang2024safe}, Jang et al. propose Lyapunov-based CBF constraints for a QP formulation. Thus, variations of CLF-CBF-RL-QP are observed in the literature to guarantee stability and satisfy state constraints. Mandal et al. \cite{mandal2024formally} propose a framework that verifies neural Lyapunov barrier (NLB) certificates using a counterexample-guided interactive synthesis (CEGIS) loop \cite{katz2019marabou} to obtain the RL controller. CEGIS loop augments the training dataset in case the policy does not satisfy the constraints of the certificate. Reach-while-avoid certificates are used as the NLBs with further extensions proposed in terms of filtering and compositions. The extensions essentially filter the state space according to safe and unsafe regions, and compositions help in scalability of filtering based approach by parallel training of multiple controllers which satisfy safe operation and selecting a safe controller at a given state.  

\begin{sidewaystable}[htpb]
\centering
\small
\caption{Classification of combined Lyapunov--barrier (CLF/CBF/CLBF) approaches in safe RL.}
\label{tab:lyap_barrier_adv_disadv}
\setlength{\tabcolsep}{4pt}
\renewcommand{\arraystretch}{1.2}
\begin{tabular}{@{}p{0.15\linewidth} p{0.33\linewidth} p{0.24\linewidth} p{0.24\linewidth}@{}}
\toprule
\textbf{Type} & \textbf{Methodology} & \textbf{Pros} & \textbf{Cons} \\
\midrule

CLF--CBF--QP filter (incl.\ temporal-logic, uncertainty-aware, CDBF/TCBF variants) &
Nominal $u_{RL}$ corrected by a QP with CLF and CBF constraints, relaxed by $\delta$; variants add FSA-based task specification, learned uncertainty terms, or richer CBF classes (control-dependent, time-varying) \cite{li2019temporal,choi2020reinforcement,meng2023reinforcement,jang2024safe} &
Stability and safety in one layer; plug-and-play with PPO/DDPG/SAC; flexible for task, uncertainty, or time-varying constraints; prevents catastrophic rollouts when feasible &
Slack weakens guarantees; per-step QP cost; model and Lie-derivative dependence; action filtering biases gradients and off-policy credit; richer variants $\Rightarrow$ harder tuning \\

CLBF / Lyapunov--barrier critic (ADP, backstepping, recentered barrier) &
Single Lyapunov--barrier object (e.g.,\ $Q_{LB}$, or recentered $H(x)=(h(x)-h(0))^2$) used as the critic; Lagrangian enforces CMDP constraint costs; backstepping handles nonlinearities and derivative terms \cite{9827972,wei2024asymmetric,du2023reinforcement,cohen2023safe,wills2004barrier} &
No per-step QP; end-to-end learning with safety in critic; CMDP-compatible and data-driven; interpretable stability + constraint structure &
Conditions hard to satisfy under function approximation or distribution shift; multiplier tuning brittle; model and derivative dependence; constraints often only approximate unless paired with a filter \\

Actor--critic with backup policies or NODE dynamics &
Actor--critic losses augmented with Lyapunov + barrier constraints; backup controller activated if learned policy is unsafe; NODE predicts continuous dynamics for constraint evaluation \cite{zhao2023stable,zhao2024nlbac,zhao2025nlbac,cocaul2024safe} &
Practical safety via fallback; NODE reduces dynamics mismatch; flexible in deep RL; risk terms shape learning &
Safety hinges on certificate, dynamics, and backup quality; more components $\Rightarrow$ more failure modes; weight/multiplier tuning sensitive \\

Verification-driven neural Lyapunov-barrier (NLB) certificates &
Train policy and neural certificates with formal verification (CEGIS); counterexamples augment training data until certificate constraints hold \cite{mandal2024formally} &
Stronger assurance than empirical training; counterexamples expose failure modes; supports reach-while-avoid certificates &
High verification cost; CEGIS hard to scale to high-dimensional control; guarantees depend on verifier and certificate class; many iterations may be needed \\

Lyapunov--barrier reward shaping (emerging) &
CLF/CBF terms added to rewards to guide learning, including diffusion-model guided control \cite{mizuta2024cobl,wang2024control} &
Easy to add; encourages safe and stable behavior; can improve exploration and sample efficiency &
Soft safety only (no invariance); shaping can distort optimality; limited hard-safety evidence; theory for joint CLF/CBF shaping remains sparse \\

\bottomrule
\end{tabular}
\end{sidewaystable}

Compared to prior approaches where a CLF-CBF-QP is solved to compensate for RL control output, Zhang et al.~\cite{9827972} propose a Barrier Lyapunov function-based RL approach for tracking and planning of vehicle motion control using an ADP algorithm. The unknown optimal policy is reformulated using Barrier-Lyapunov functions method into model-based and adaptive parts. The actor-critic networks are used to approximate the control policy and value function. Backstepping is employed to achieve global controllability of the system by optimizing virtual control in each subsystem. As in the case of prior approaches, backstepping is introduced to approximate the time derivative of $V$ along the dynamics for CLF and CBF. Du et al.~\cite{du2023reinforcement} propose a purely data-driven approach using a control Lyapunov Barrier function in an actor-critic formulation for 2D quadrotor navigation. The control Lyapunov Barrier function (CLBF) is chosen as the critic function in a CMDP formulation with constraints based on data-based CLBF theorem (Section IV-B of cited work). The constrained optimization problem is augmented to an unconstrained problem using Lagrange multipliers and updated through stochastic gradient descent as follows,
\begin{equation}
	\begin{aligned}
		J(\phi) = \EX_{\mathcal{D}} \bigg[ \frac{1}{2} (Q_{\text{LB}}(x,u) - Q_{\text{target}}(x,u))^2 \\
		+\lambda \bigg(Q_{\text{LB}}(x^{\prime}, \pi_{\theta}(x^{\prime}))\mathbbm{1}_{\Delta}(x^{\prime}) \\
		- Q_{\text{LB}}(x, \pi_{\theta}(x))\mathbbm{1}_{\Delta}(x) + \alpha d\bigg)\bigg]
	\end{aligned}	
\end{equation}
where, $Q_{\text{LB}}$ is the Lyapunov-barrier function, $\mathbbm{1}_{\Delta}(\cdot)$ denotes the indicator function with respect to state $x$ being in the safe set ($\Delta = \mathcal{X} (\mathcal{X}_{\text{goal}} \cup \mathcal{X}_{\text{unsafe}})$), $\lambda$ denotes the Lagrange multiplier, $\alpha$ is a positive constant and $d$ denotes the upper bound on constraint cost in a CMDP. Thus, the proposed $Q_{\text{LB}}$ satisfies the system and state constraints by satisfying the CMDP cost constraint at each step. Cohen et al.~\cite{cohen2023safe} follow a similar approach of using a Lyapunov-like CBF for a collision avoidance application in a model-based ADP formulation. The Lyapunov-like CBF is expressed as a re-centered barrier function~\cite{wills2004barrier},
\begin{equation}
	H(x) = (h(x) - h(0))^2
\end{equation}
which gives suitable Lyapunov function properties of positive semi-definiteness. The barrier function is re-centered at the origin, thus, the barrier function vanishes at the origin. The proposed Lyapunov CBFs are added as a compensating controller (similar to \ref{eqn:CBF_QP_u}) in the control policy to drive the control input to safe sets and guarantee stability. The problem is formulated as an ADP and tested on collision avoidance and single integrator system with unknown drift dynamics. Zhao et al.~\cite{zhao2023stable} propose a barrier-Lyapunov actor-critic approach using a Lagrangian formulation for the actor and critic loss functions. A backup controller is proposed in case valid control input cannot be computed by the actor-critic RL controller ensuring safety and stability constraints. The authors also propose in \cite{zhao2024nlbac} a neural ordinary differential equation (NODE)-based framework where NODE is used to predict the system dynamics. The remaining framework remains similar to their prior work using the barrier-Lyapunov actor-critic approach. Authors in \cite{cocaul2024safe, zhao2025nlbac} follow a similar methodology with \cite{cocaul2024safe} incorporating the empirical Lyapunov risk term in the critic loss function.

Few approaches consider Lyapunov-barrier-based reward-shaping. Notably, Mizuta et al. \cite{mizuta2024cobl} propose CLF and CBF terms to guide the denoising process of a diffusion model. CLF and CBF reward functions are proposed for a single- and multi-agent pedestrian datasets. Since we focus on safe RL in this review, a detailed explanation is omitted and the idea to introduce this work is to provide possible research directions for using reward-shaping. 
%%%%%%%%%%%%%%%%%%%%%%%%%%%%%%%%%%%%%%%%%%%%%%%%%%%%%%%%%%%%%%%%%%%%%%%%%%%%%%%%%%%%%%%%%%%%%%%%%%%%%%%%%%%%%%%%%%%
\subsection{Classification and Discussion}
\label{subsection:ControlCLBF}

Table \ref{tab:lyap_barrier_adv_disadv} provides a tabular comparison of each methodology and its shortcomings. The surveyed literature can be broadly grouped into three recurring templates: (i) \emph{CLF-CBF constrained optimization filters}, where an RL action is compensated via a per-step quadratic program (QP) constrained by both Lyapunov and barrier conditions; (ii) \emph{Lyapunov-barrier critics and Lagrangian learning}, where a single Lyapunov-barrier function (e.g., CLBF, Lyapunov-like CBF) is embedded into the critic or actor-critic objective; and (iii) \emph{Lyapunov--barrier reward shaping}, where CLF/CBF terms shape rewards to bias exploration toward stability and constraint satisfaction.

\paragraph{CLF-CBF constrained optimization filters}
A dominant approach is to treat the RL controller as a nominal policy $u_{RL}$ and compute the final action using a CLF--CBF-QP safety filter. In~\cite{li2019temporal}, temporal-logic-guided RL augments the MDP with a finite-state automaton and solves a relaxed CLF--CBF-QP to prevent deadlock between stabilization and constraint satisfaction. The resulting control law is of the form $u_k(x)=u_k^{RL}+u_k^{\text{CLF-CBF}}(x,u_k^{RL})$, where the compensating term is obtained by enforcing Lyapunov decrease and barrier constraints. Variants of this paradigm also appear in~\cite{choi2020reinforcement,meng2023reinforcement,jang2024safe}, differing primarily in (i) how uncertainty/disturbances are learned or estimated (e.g., learning uncertainty terms in constraints \cite{choi2020reinforcement}), and (ii) the choice of barrier type (e.g., control-dependent and time-varying CBFs \cite{meng2023reinforcement}) or Lyapunov-based barrier formulations \cite{jang2024safe}. While these filters offer strong practical safety when feasible, they raise important questions regarding feasibility under simultaneous CLF and CBF constraints, the computational cost of solving a constrained optimization at each time step, and the induced conservatism that can limit exploration.

\paragraph{Lyapunov-barrier critics and Lagrangian formulations}
A second class of methods integrates Lyapunov--barrier structure directly into the learning objective to avoid (or complement) per-step QPs. Zhang et al.~\cite{9827972} employ barrier-Lyapunov formulations in an ADP setting for vehicle tracking, using actor--critic approximations and backstepping to handle nonlinearities and derivative terms. Du et al.~\cite{du2023reinforcement} propose a data-driven control Lyapunov--barrier function (CLBF) as the critic in a CMDP-based actor--critic formulation, using Lagrangian multipliers to enforce constraint costs on the learned Lyapunov--barrier critic. Cohen et al.~\cite{cohen2023safe} similarly leverage a Lyapunov-like CBF via a re-centered barrier construction to obtain positive semi-definite Lyapunov properties, embedding the resulting certificate in an ADP framework for collision avoidance. Zhao et al.~\cite{zhao2023stable} propose a barrier-Lyapunov actor--critic method with Lagrangian losses and incorporate a backup controller when the learned policy cannot provide valid safe actions; follow-up work introduces NODE-based dynamics prediction to support the same learning framework \cite{zhao2024nlbac}, with related extensions including empirical Lyapunov risk in critic learning \cite{cocaul2024safe,zhao2025nlbac}. These methods reduce reliance on online QP solving but shift the main challenge to reliably learning certificates from sampled data and ensuring generalization beyond the training distribution.

\paragraph{Formal verification and certificate synthesis}
Mandal et al.~\cite{mandal2024formally} propose verifying neural Lyapunov barrier (NLB) certificates using a counterexample-guided interactive synthesis (CEGIS) loop \cite{katz2019marabou}, augmenting the training set whenever certificate constraints are violated. Such verification-oriented pipelines strengthen assurance but add computational overhead and face scalability challenges when certificates, neural policies, or state dimensions grow.

\paragraph{Lyapunov--barrier reward shaping}
Finally, a smaller body of work explores shaping rewards using both CLF and CBF terms. Mizuta et al.~\cite{mizuta2024cobl} incorporate CLF/CBF rewards to guide diffusion-model denoising, illustrating a broader design space for reward shaping even though it is not purely within standard safe RL control benchmarks. In general, Lyapunov--barrier shaping remains comparatively under-explored in continuous-control safe RL, especially with hard safety objectives.
%%%%%%%%%%%%%%%%%%%%%%%%%%%%%%%%%%%%%%%%%%%%%%%%%%%%%%%%%%%%%%%%%%%%%%%%%%%%%%%%%%%%%%%%%%%%%%%%%%%%%%%%%%%%%%%%%%%
\subsection{Key Issues and Open Problems}
\label{subsec:clbf_open}

\paragraph{Feasibility and complexity of joint CLF--CBF constraints}
Simultaneously enforcing Lyapunov decrease and barrier invariance can create infeasible QPs or induce conservative solutions (especially near boundaries), motivating adaptive relaxation strategies, feasibility certificates, and principled slack allocation \cite{li2019temporal,meng2023reinforcement}.

\paragraph{Learning certificates from data and guaranteeing generalization}
Model-free CLBF/Lyapunov-like CBF critics rely on sampled transitions to learn certificates \cite{du2023reinforcement,cohen2023safe}. Establishing conditions under which learned certificates remain valid under approximation error and distribution shift is still an open problem.

\paragraph{Closed-loop interactions between optimization filters and RL updates}
When per-step QPs are used, the executed action differs from $u_{RL}$, which can bias gradients and off-policy learning, and complicate credit assignment. Differentiable QP layers and stability-aware policy gradient theory for filtered policies remain important directions.

\paragraph{Scalable verification and synthesis of neural certificates}
Verification-driven pipelines (e.g., CEGIS) can significantly improve assurance \cite{mandal2024formally}, but scaling verification and counterexample generation to high-dimensional continuous control remains challenging.

\paragraph{Unified design principles for CLBF selection}
Across works, CLF/CBF choices are often problem-specific. Developing systematic CLBF parameterizations (e.g., structured neural certificates, compositional certificates, reach-while-avoid certificates) and practical guidelines for selecting Lyapunov and barrier candidates is a key research opportunity.

Table~\ref{tab:CBF_CLF} summarizes the literature reviewed for cumulative use of Lyapunov and barrier functions in RL. Overall, combined Lyapunov--barrier methods offer a compelling route to jointly guarantee stability and constraint satisfaction, but their broad deployment hinges on resolving feasibility/computation barriers for constrained optimization filters and establishing robust learning and verification principles for data-driven certificates.

\subsection{Comparative Analysis}\label{subsec:cross_cutting_comparison}
Synthesizing across Sections~\ref{section:ControlLF}--\ref{section:CLFandCBF}, the three method families differ along five orthogonal axes:
\begin{enumerate}
  \item \emph{Safety enforcement strength.} Combined CLF--CBF--QP filters provide the strongest practical safety when feasible (hard invariance + asymptotic stability); pure CBF--QP filters provide hard invariance only; CMDP--Lagrangian methods provide expectation-level safety; reward shaping provides only soft safety.
  \item \emph{Model dependence.} Combined and CBF--QP methods require control-affine structure and Lie derivatives. Lyapunov-as-value methods can dispense with explicit dynamics if the critic is learned, but at the cost of weaker certificate validity.
  \item \emph{Sample efficiency.} CBF--QP shielding $>$ neural certificate methods $>$ Lyapunov-as-value $>$ CMDP--Lagrangian (Section \ref{subsec:data_efficiency}).
  \item \emph{Scalability.} Reward shaping and Lagrangian methods scale best with state dimension; combined CLF--CBF--QP methods scale worst due to joint feasibility constraints (Section \ref{subsection:Current_Challenges}).
  \item \emph{Exploration vs.\ exploitation.} Hard-shielded methods restrict exploration the most; reward shaping the least.
\end{enumerate}
A direct quantitative head-to-head comparison across reviewed works is not generally possible because benchmarks (Cart-Pole, MuJoCo, Safe-Control-Gym, custom robotics simulators), constraint specifications, and reported metrics (cumulative reward vs.\ violation rate vs.\ sample budget) differ substantially. Where benchmarks overlap, most commonly inverted pendulum, Cart-Pole, and 2D quadrotor where reported safety-violation rates for hard-shielded methods are typically zero or near-zero during deployment~\cite{cheng2019end,emam2021safe,cai2021safe}, while reward-shaping methods report violation rates in the 1--10\% range~\cite{marvi2021safe,dong2020principled}. We present this synthesis as guidance rather than a leaderboard, and recommend that future work in this area adopt Safe-Control-Gym (Section \ref{subsec:Benchmarks}) to enable apples-to-apples comparison.

%%%%%%%%%%%%%%%%%%%%%%%%%%%%%%%%%%%%%%%%%%%%%%%%%%%%%%%%%%%%%%%%%%%%%%%%%%%%%%%%%%%%%%%%%%%%%%%%%%%%%%%%%%%%%%%%%%%
\begin{table}[htpb]
\centering
\caption{Summary of safe RL with Lyapunov/barrier certificates across the three method families reviewed.}
\label{tab:unified_summary}
\small
\setlength{\tabcolsep}{4pt}
\renewcommand{\arraystretch}{1.25}
\begin{tabularx}{\textwidth}{@{}p{0.13\textwidth} X X X X@{}}
\toprule
\textbf{Family} & \textbf{Safety mechanism} & \textbf{Model dependence} & \textbf{Guarantee type} & \textbf{Scalability bottleneck} \\
\midrule
Lyapunov-only (\S\ref{section:ControlLF}) & Stability via attractive level sets; reward shaping; ROA expansion; CMDP feasibility & Low (model-free) to High (ADP/HJB) & Asymptotic; recently finite/fixed-time~\cite{tan2025finite, tan2025finite} & Discretized verification: $n \lesssim 8$; neural certificates: $n \lesssim 100$ \\
Barrier-only (\S\ref{section:ControlBF}) & Forward invariance via CBF--QP shielding; reward shaping; Lagrangian; bilevel & High (Lie derivatives) to Medium (GP/observers) & Hard (when QP feasible); probabilistic with slack & QP feasibility/deadlock; constraint count $p$ dominates \\
Combined CLF--CBF (\S\ref{section:CLFandCBF}) & Joint stability + invariance via CLF--CBF--QP; CLBF critics; verification (CEGIS) & High (control-affine) to Medium (NODE/data-driven) & Hard + asymptotic when feasible; verified for NLBs & Joint feasibility under model error; CEGIS scaling \\
\bottomrule
\end{tabularx}
\end{table}

\begin{figure}[htpb]
\centering
\begin{tikzpicture}
\begin{axis}[
  ybar stacked,
  width=0.95\linewidth, height=6cm,
  bar width=8pt,
  ymin=0, ymax=20,
  ylabel={Number of reviewed papers},
  xlabel={Publication year},
  symbolic x coords={2002,2015,2016,2017,2018,2019,2020,2021,2022,2023,2024,2025,2026},
  xtick=data, x tick label style={rotate=45,anchor=east},
  legend style={at={(0.02,0.98)},anchor=north west,legend columns=1,font=\small},
  enlarge x limits=0.05
]
\addplot+[fill=blue!50] coordinates {
  (2002,1)(2015,1)(2016,1)(2017,2)(2018,1)(2019,1)(2020,4)(2021,2)(2022,1)(2023,4)(2024,4)(2025,8)(2026,0)
}; \addlegendentry{Lyapunov-only}
\addplot+[fill=red!50] coordinates {
  (2002,0)(2015,0)(2016,0)(2017,0)(2018,0)(2019,2)(2020,0)(2021,4)(2022,6)(2023,6)(2024,8)(2025,2)(2026,0)
}; \addlegendentry{Barrier-only}
\addplot+[fill=green!50] coordinates {
  (2002,0)(2015,0)(2016,0)(2017,0)(2018,0)(2019,1)(2020,1)(2021,0)(2022,1)(2023,4)(2024,8)(2025,2)(2026,2)
}; \addlegendentry{Combined CLF--CBF}
\end{axis}
\end{tikzpicture}
\caption{Distribution of reviewed papers by publication year and methodology family. The post-2022 surge motivates this review's focus on recent literature; combined CLF--CBF methods only become substantial after 2020.}
\label{fig:lit_trends}
\end{figure}
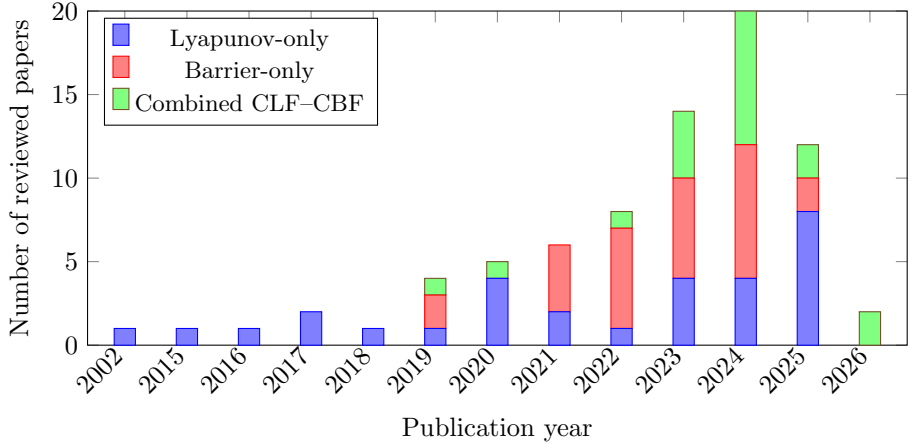

\section{Discussion}\label{section:Discussions}
Safe RL with Lyapunov or barrier functions is an effective and promising field that aims to ensure the safety of an agent while it learns a task through RL. Tables~\ref{tab:CLF},~\ref{tab:CBF}, and~\ref{tab:CBF_CLF} summarize the methodology of using Lyapunov and barrier functions in RL. To make the qualitative shift in the literature more explicit, Fig.~\ref{fig:lit_trends} summarizes the approximate distribution of the reviewed certificate-based safe RL papers by time period and certificate type. The counts are based on the papers included in Tables~\ref{tab:CLF},~\ref{tab:CBF}, and~\ref{tab:CBF_CLF}; they are intended as a descriptive map of this review's coverage rather than a full bibliometric census.

\subsection{Simulation Benchmarks}\label{subsec:Benchmarks}
Simulated benchmarks have proven to be a norm in implementing and validating data-driven ML and RL algorithms over the last few decades\cite{dulac2021challenges}. The key challenges involved in creating a successful benchmark for these data-driven algorithms are reproducibility, ease of implementation, computational complexity (should be manageable to allow testing on all scales of hardware), incorporating stochastic behavior and sufficient environment complexity (indicating the environment goals should be sufficiently complex). We review some of the existing environments for benchmarking safe RL and address its usefulness in validating methodologies incorporating Lyapunov and barrier functions. 

\begin{figure}
	\centering
	\begin{subfigure}{0.7\textwidth}
		\centering
		\includegraphics[]{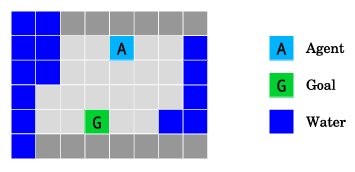}
		\label{fig:GridWorld}
		\caption{(A) Island Navigation environment \cite{leike2017aisafetygridworlds}}
	\end{subfigure}
	\begin{subfigure}{0.7\textwidth}
		\centering
		\includegraphics[]{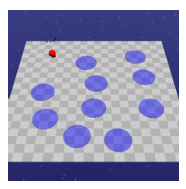}
		\label{fig:SafeGym}
		\caption{(B) Constraints in Safe Gym \cite{ray2019benchmarking}}
	\end{subfigure}
	\caption{Environments: Gridworlds and Safe Gym}
\end{figure}

The earliest benchmark for safe RL approached the problem as a safe exploration towards a goal. DeepMind introduced AI Safety Gridworlds\footnote{\url{https://github.com/google-deepmind/ai-safety-gridworlds}} \cite{leike2017aisafetygridworlds} and OpenAI released Safety Gym\footnote{\url{https://github.com/openai/safety-gym}} \cite{ray2019benchmarking} as benchmarks for safe exploration problem in RL. Safety Gym proposes a constrained RL formulation for safe exploration and provides gym\footnote{\url{https://github.com/openai/gym}} \cite{brockman2016openai} environments for high-dimensional continuous control in safe RL. Safety Gym was upgraded by introducing Safety-Gymnasium\footnote{\url{https://github.com/PKU-Alignment/safety-gymnasium}} \cite{ji2023safety} with increased functionality by using MuJoCo physics engine\footnote{\url{https://github.com/google-deepmind/mujoco}} \cite{todorov2012mujoco}, higher customization, extended single and multi-agent scenarios. 
Safety-Gymnasium is built on Gymnasium\footnote{\url{https://github.com/Farama-Foundation/Gymnasium}} \cite{towers2024gymnasium} RL environment structure and allows for easy integration with RL libraries. Safety-Gymnasium also proposes SafePO which provides single and multi-agent approaches for constrained safe RL problems. The ideal benchmarking environment for reviewed methodologies should incorporate constraints, external disturbances, nonlinear dynamical models and low-level oriented tasks. For this reason Safe-Control-Gym\footnote{\url{https://github.com/utiasDSL/safe-control-gym}} \cite{9849119} appears as the most suitable benchmark environment in the authors' perspective. Safe-Control-Gym provides environments for three nonlinear dynamical systems: (i) Cart-Pole; (ii) 1D and 2D Quadrotor; (iii) 3D Quadrotor. The quadrotor environments have the following tasks:
\begin{itemize}
	\item[1.] Stabilization - hovering quadrotor at a fixed point and Cart-Pole stabilization.
	\item[2.] Trajectory Tracking - tracking a predefined trajectory. Includes functionality to introduce custom trajectories.
\end{itemize}
The environment uses PyBullet physics engine\footnote{\url{https://github.com/bulletphysics/bullet3}} \cite{coumans2019} to incorporate dynamic disturbances and external forces applied on the agent during simulation. These can be customized using the environment API to introduce necessary exogenous inputs and simulate uncertainty in the system. The environment API also allows for linear and quadratic custom constraints to be added to the states and control input. Overall, the environment provides all the necessary aspects to test and validate proposed safe RL methodologies.

Although MuJoCo~\cite{todorov2012mujoco} offers high-fidelity rigid-body dynamics, it is not natively a safe-RL benchmark: it does not expose a built-in API for symbolic constraint specification, does not provide a structured mechanism for injecting parametric or process disturbances, and does not include reference safe controllers (LQR, MPC, PID) for fair comparison. Safe-Control-Gym addresses these by exposing (i) symbolic models of the dynamics that allow analytic Lie derivatives required by CBF/CLF formulations, (ii) a constraint API for linear and quadratic state/input bounds, and (iii) disturbance and uncertainty injection at the parameter, state, and observation levels. These features make it the most suitable environment for the methods reviewed here, where certificate computation typically requires the symbolic dynamics that MuJoCo abstracts away.

There are several third-party and open--source environments for safe RL, however, their reliability and reproducibility have been under scrutiny in the safe RL community. For the purpose of this review, we focus on peer-reviewed, tested and validated environments in the literature.

\subsection{Theoretical developments}\label{subsection:Theoretical_dev}
Some key observations from the table suggest a shift from model-based to model-free approaches in more recent years. The theoretical developments based on literature reviewed are as follows:

\paragraph{Formal safety guarantees}
One of the main advantages is the ability to provide formal safety guarantees. This provides a level of confidence that the agent will not deviate into unsafe states during learning or deployment. It is observed through this review that most model-based approaches provide strict guarantees compared to probabilistic or confidence bounds provided by model-free RL approaches. This is expected as the model-based RL has prior information about the dynamic structure of the system~\cite{vamvoudakis2015asymptotically, kamalapurkar2016model, chang2021stabilizing}. 

\paragraph{Risk mitigation during training}
Lyapunov and barrier functions provide theoretical guarantees for the agent to actively avoid unsafe states or actions during the learning process as well. This proactive safety mechanism reduces system constraint violation (the risk of accidents or failures) during exploration while training and deployment, making them well-suited for applications in safety-critical domains~\cite{garcia2015comprehensive, brunke2022safe}. These approaches can be useful in deploying agents during the training process as well.  

\paragraph{Adaptive safety and safe exploration}
These methods can adapt to changes in the environment and system dynamics. Given most applications involve stochastic environments, the agent should be able to adjust its behavior to accommodate variations and unforeseen events \cite{cheng2023safe}. In~\cite{cai2021safe}, authors incorporate CBFs in multi-agent RL for cooperative and non-cooperative agents to achieve $0\%$ collision for a two man patrol problem. Safe exploration is a challenge in RL, especially in environments with high risks of failure leading to complete breakdown of the system. Lyapunov and barrier functions can guide exploration by constraining the agent's actions within safe bounds. This ensures that an agent safely explores the environment by enforcing hard constraints (QP-based filters \cite{hou2024safe, liu2025safe}), providing formal guarantees during training and deployment~\cite{marvi2021safe, emam2021safe}.

\paragraph{Smooth transitions (magnitude of control input) to safe states}
Lyapunov-barrier functions can enable smooth transitions of the agent from unsafe to safe states by tuning the accepted risk parameter and formulating a minimum control magnitude QP problem. This is particularly useful in scenarios where abrupt actions needed to avoid unsafe states could themselves lead to instability or undesired actions~\cite{10155918, marvi2021safe}.

\begin{figure*}[!t]
	\centering
	\includegraphics[width=0.95\textwidth]{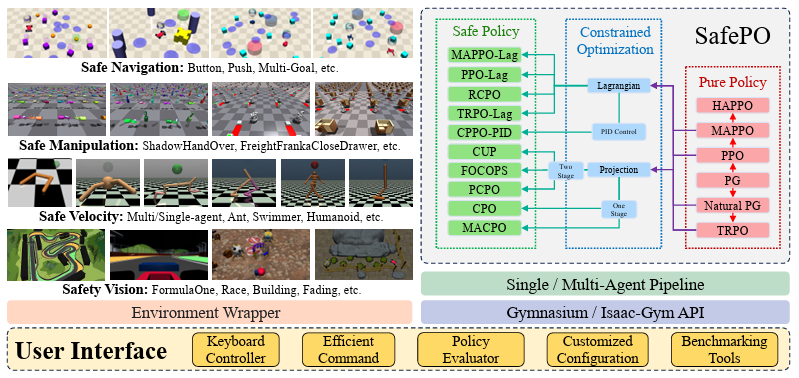}
	\caption{Safe Policy Optimization environment architecture \cite{ji2023safety}}
	\label{fig:SafePO}
\end{figure*}

\subsection{Convergence Properties}\label{subsection:convergence}
Convergence guarantees in safe RL with Lyapunov/barrier certificates have evolved along three regimes. \emph{Asymptotic} convergence---trajectories approach the equilibrium or safe set as $t \to \infty$---is established for most ADP-style works~\cite{vamvoudakis2015asymptotically,kamalapurkar2016model} under persistence-of-excitation and Lipschitz assumptions, with convergence rates that depend on excitation gain and approximation error bounds. \emph{Finite-time} convergence provides explicit settling-time bounds that depend on initial conditions, derived through homogeneity-based Lyapunov analysis or terminal sliding-mode-style certificates. Recent work~\cite{tan2025finite} extends safe RL to multi-player nonzero-sum games for quadcopter systems with finite-time guarantees, addressing both stability and safety through a coupled CLF--CBF formulation. \emph{Fixed-time} convergence provides settling-time bounds independent of initial conditions, a stronger property particularly valuable in safety-critical applications where worst-case settling time must be apriori bounded; \cite{tan2025fixed} establishes fixed-time convergence for stochastic learning from human--UAV interaction under state constraints. These advances bound time-to-stability spatially via the Lyapunov ROA and CBF safe set, and temporally via finite/fixed-time analysis, providing complementary deployment guarantees. The trade-off is that finite/fixed-time results typically require fractional-power Lyapunov derivatives, which complicate function approximation with smooth neural networks and can introduce chattering near the equilibrium.

\subsection{Soft vs.\ Hard Safety Constraints}\label{subsec:soft_vs_hard}
The reviewed approaches occupy a continuum from \emph{hard} (strict forward invariance, e.g., feasible CBF--QP shielding~\cite{cheng2019end,emam2021safe}) to \emph{soft} (constraint-violating with bounded penalty, e.g., reward shaping~\cite{marvi2021safe,dong2020principled} or Lagrangian methods~\cite{chow2018lyapunov,zhang2024constrained}). Hard safety is appropriate when even a single violation incurs catastrophic cost such as collisions in autonomous driving~\cite{cheng2019end}, attitude limits in aerospace~\cite{chen2026safe}, voltage limits in power systems~\cite{vu2021barrier}. But these require accurate models, feasible QPs, and tolerate conservatism. Soft safety is appropriate during exploration in simulation, when constraints encode preferences rather than physical limits, or when soft violations can be recovered from (e.g., minor velocity overshoot). The slack-relaxed CBF--QP~\cite{wang2022ensuring} is a hybrid: hard safety when feasible, soft safety with state-dependent penalty otherwise. A practical guideline emerging from this review is to pair hard execution-time filters with soft training-time shaping, combining sample efficiency during learning with deployment-grade safety.

\subsection{Data Efficiency}\label{subsec:data_efficiency}
Sample efficiency varies substantially across the reviewed approaches. CBF--QP shielding methods~\cite{cheng2019end,emam2021safe} typically require fewer samples to reach safety thresholds because unsafe actions are filtered before execution, preventing wasteful unsafe rollouts; reported sample reductions on inverted pendulum and car-following are 30--60\% versus unshielded baselines. Reward-shaping methods~\cite{marvi2021safe,dong2020principled} accelerate convergence through guided exploration but provide weaker sample-efficiency gains (typically 10--25\%). CMDP--Lagrangian methods~\cite{chow2018lyapunov,zhang2024constrained} are generally less sample-efficient due to dual-variable updates that can oscillate; sample efficiency improves with PID-controlled Lagrangian updates~\cite{ji2023safety}. Neural certificate methods~\cite{chang2021stabilizing,mandal2024formally} sit in the middle: they amortize cost across episodes by learning a single certificate that informs all subsequent updates. The empirical pattern is that hard execution-time filters dominate on sample efficiency for reaching safety thresholds, while differentiable certificates dominate for asymptotic performance under fixed sample budgets.

\subsection{Recent Extensions: SOS Verification, Continuous-Time Safe RL, Offline Safe RL}\label{subsec:recent_extensions}
Three rapidly developing directions complement the core Lyapunov/barrier RL literature reviewed above.
\paragraph{Sum-of-squares verification.} SOS programming provides automated certificate synthesis when both the dynamics and the candidate certificate are polynomial~\cite{huang2022barrier,kang2022lyapunov}. Coupled with neural-policy approximation, SOS converts certificate-checking into a semi-definite program and is complementary to CEGIS-based verification used in NLB pipelines~\cite{mandal2024formally}.
\paragraph{Continuous-time safe RL.} Recent work extends safe RL beyond the discrete-time MDP formulation to continuous-time dynamics with state constraints and bounded disturbances~\cite{zhang2023robust}, providing stronger semantics for forward-invariance arguments at the cost of requiring continuous-time data or careful sampled-data discretization analysis. Neural ODE-based critics~\cite{zhao2024nlbac} are a hybrid that retains discrete-time RL updates while modeling continuous-time dynamics in the critic.
\paragraph{Offline safe RL.} Offline safe RL targets safety under purely batch data without environment interaction; constraint-penalized Q-learning~\cite{xu2022constraints} and behavior-regularized methods~\cite{le2019batch} enforce constraint costs through penalty terms or distributional support constraints. Combining offline learning with Lyapunov/barrier certificates (e.g., distributionally robust certificate validation under offline-to-online distribution shift) is a largely open direction.

\subsection{Current Challenges}\label{subsection:Current_Challenges}
While these approaches provide strong theoretical guarantees and improvement in terms of performance metrics and convergence rates, several open questions and challenges remain as discussed below: 

\paragraph{Conservatism and performance trade-off (exploration vs exploitation)}
RL primarily relies on exploring the environment to find an optimal (or effective) control policy. A key challenge in using Lyapunov or barrier functions in safe RL is the conservative exploration introduced by these safety constraints on the derived policy, often leading to suboptimal results. These constraints are designed to ensure forward invariance of the system  within a safe set. This results in constrained or overly cautious/conservative exploration, limiting the agent's ability to explore and learn efficiently. This creates a trade-off between safety and optimality resulting in a conundrum for the agent in achieving its objective safely. There have been several approaches to address this issue by perturbing the safe set~\cite{westenbroek2022lyapunov, ohnishi2019barrier, chow2019lyapunov}. However, most approaches present a tunable hyperparameter to adjust the allowable safe set perturbation. Formal or probabilistic bounds for the magnitude of perturbation for the safe set are an open area for research.

\paragraph{Complexity and computational costs for systems with high dimensionality}
Designing or approximating Lyapunov and/or barrier functions that accurately capture the safety requirements of a complex environment can be challenging since they require some prior information about the dynamic structure of the model. As the complexity of the environment increases, finding appropriate Lyapunov or barrier functions becomes even more difficult. Furthermore, most approaches incorporate safety constraints as a constrained optimization problem to compute the safe action at each iteration which increases complexity as dimensionality of the problem increases~\cite{westenbroek2020learning, ma2021model,cai2021safe,ohnishi2019barrier}.

We summarize concrete complexity bounds for the dominant computational primitives in safe RL with Lyapunov/barrier certificates. For a quadratic program with $m$ decision variables and $p$ inequality constraints, dense interior-point solvers attain worst-case complexity $\mathcal{O}((m+p)^3)$ per solve, while operator-splitting solvers such as OSQP~\cite{stellato2018osqp} and active-set methods such as qpOASES~\cite{ferreau2014qpoases} achieve $\mathcal{O}((m+p)^2)$ per iteration with a small problem-dependent number of iterations. For a control-affine system of state dimension $n$ and input dimension $m_u$, a single CBF--QP filter typically uses $m = m_u + 1$ (control plus slack) and $p = O(n_c)$ where $n_c$ is the number of barrier conditions; joint CLF--CBF filters double the constraint count, scaling primarily with $p$. Discretized Lyapunov-decrease verification over a uniform grid of resolution $\tau$ scales as $\mathcal{O}(\tau^{-n})$ in state dimension $n$, exposing the curse of dimensionality. Policy-network forward passes scale as $\mathcal{O}(L h^2)$ for $L$ layers of width $h$, and CEGIS verification loops add an outer factor proportional to the number of counterexample-augmentation rounds, with each round dominated by a SMT/Marabou-style query whose worst-case cost is exponential in network width.

To mitigate per-step QP cost, recent work distills the QP solution into a feedforward neural network, replacing the optimization with an $\mathcal{O}(L h^2)$ forward pass. OptNet~\cite{amos2017optnet} and CBF-specialized differentiable QP layers admit gradients through the safety layer, enabling end-to-end training. Explicit-MPC-style precomputation can further amortize CBF--QP solving when the constraint set is fixed.

Empirically, the practical scaling limits observed across the reviewed literature are: discretized Lyapunov-decrease verification (e.g., \cite{berkenkamp2017safe}) is feasible up to $n \approx 6$--$8$ before the curse of dimensionality dominates; CBF--QP filters with hand-designed certificates scale to $n \approx 20$--$30$ with control-affine structure (e.g., \cite{cheng2019end}); neural-certificate approaches~\cite{chang2021stabilizing,mandal2024formally} have been demonstrated up to $n \approx 50$--$100$ on benchmark continuous-control tasks; multi-agent decentralized CBF shields~\cite{cai2021safe} scale to roughly 10--20 agents in 2D/3D environments before coupled-feasibility issues dominate. CEGIS-based verification remains the most dimensionally restrictive, typically limited to $n \leq 12$.

\paragraph{Limited applicability for dynamic and unknown environments}
Using Lyapunov and barrier functions in safe RL produces several challenges in case of partially observable environments. They often assume a certain level of structure or regularity in dynamics of the system, which might not hold in highly unpredictable or stochastic and partially observable environments. This is an open area of research given its applicability to real-world problems. This limitation is discussed as the drawback of using model-free RL approaches. How much data is sufficient to get an accurate approximation of CLF and CBF for the true system to provide formal guarantees remains an open question. Furthermore, generalizing safety constraints to unknown and partially observable environments is scarcely present in the literature. If the learned safety constraints are too tailored to the training environment, they might not transfer effectively to different contexts, leading to unsafe behavior when the agent encounters unknown scenarios.

\paragraph{Current lack of real-world robustness for unmodeled dynamics and disturbances}
While Lyapunov and barrier functions can provide guarantees in theory, real-world systems often have unmodeled dynamics, sensor noise or imperfect measurements, and external disturbances. These can lead to violations and often infeasible solutions to the assumed safety guarantees, resulting in unsafe behavior. Model-free approaches seem promising in this regard as shown in~\cite{chow2018lyapunov, chow2019lyapunov, cheng2023safe}. 

\subsection{Future Work}\label{subsection:future_work}
Model-based approaches inherently seem suited to the incorporation of Lyapunov and barrier functions due to prior information available for the dynamic structure of the model. However, in most cases this is not possible due to incomplete model knowledge, unknown disturbances and data being available only through environment interaction. Model-free approaches present great opportunities in this regard. Theoretical gaps exist in terms of general conditions required for selecting the Lyapunov function as the value function~\cite{chang2021stabilizing,westenbroek2022lyapunov}. The general applicability of selecting a candidate Lyapunov function as a value function remains to be formalized.  Another area of research can be to quantify data requirements for sufficiently approximating the candidate Lyapunov or barrier functions. \cite{chang2021stabilizing} provides an explanation of this problem by proposing use of Almost Lyapunov conditions, however, the amount of data required or generalization error (performance in training compared to the entire state space) still remains an open area for research. Convergence guarantees while using deep or large NNs ($\geq 100$ hidden units) as function approximators are another promising area of research for RL algorithms. Using NNs of above-mentioned structure makes theoretical guarantees intractable. As observed in this review, several methods approximate the nonlinear dynamics to analytically obtain an expression for the Lie derivative employed in CLF and CBF QP formulations. Recent advances in approximating nonlinear dynamics as a linear system using Koopman theory~\cite{brunton2021modern} may be an interesting area to explore and help provide a generalized structure for this. 

\paragraph{Lyapunov function selection in deep RL} A persistent open problem is the lack of systematic guidelines for choosing a Lyapunov candidate when the value function is a deep neural network. Quadratic candidates derived from local linearizations are common but lose validity outside small neighborhoods, while learned candidates require regularization to avoid trivial solutions. Promising directions include input-convex neural networks for monotonicity guarantees, Zubov-structured parameterizations~\cite{wang2024actor}, and physics-informed candidates that respect known conservation laws.

\paragraph{Multi-agent settings} Single-agent CLF/CBF certificates do not directly extend to multi-agent settings due to non-stationary interaction dynamics and coupled constraints. Compositional certificates, where global safety follows from local agent-wise certificates plus structured coupling terms, and graph-neural-network parameterizations of multi-agent barriers are promising open directions.

\paragraph{Partial observability} Most reviewed approaches assume full state observability. Extending Lyapunov/barrier certificates to POMDPs requires belief-space safety, which couples certificate design with belief estimation under measurement uncertainty. Memory-augmented architectures (e.g., the Transformer-based formulation in~\cite{jeddi2021lyapunov}) and contraction-based observers offer promising starting points but lack rigorous safety guarantees in belief space.

\subsection{Potential and Current Applications}\label{subsection:Potential_Apps}
Lyapunov and barrier functions in safe RL have found significant application in robotics and autonomous systems, primarily ensuring collision avoidance, stability, and constraint adherence. Barrier functions integrated into RL have been effectively used for autonomous vehicles, providing formal guarantees for safety through collision avoidance, lane keeping, and speed control. For instance, CBFs guarantee safety-critical constraints, enabling vehicles to navigate complex traffic conditions safely while optimizing travel efficiency \cite{ames2019control}. Lyapunov-based methods have been leveraged in multi-agent UAV systems for trajectory planning and formation control. They provide theoretical guarantees of system stability and collision avoidance, critical for complex multi-agent coordination tasks \cite{wang2022distributed, wang2022ensuring}. Recent UAV/UGV applications have explored hierarchical safe RL architectures that combine high-level task planning with low-level safety-certified control. Adaptive hierarchical control of quadcopters via safe RL from human demonstration~\cite{tan2026adaptive} learns demonstration-aligned high-level policies while enforcing CLF/CBF safety at the actuator level, and hierarchical safe RL for leader--follower systems with prescribed performance~\cite{tan2025hierarchical} integrates prescribed-performance bounds with barrier-Lyapunov certificates for guaranteed-tracking under formation constraints. Barrier functions have been used in industrial robots to ensure safety when interacting with dynamic environments or humans. These methods ensure physical safety without compromising operational efficiency, especially in collaborative robot settings  \cite{long2025certifying}.

\begin{figure}[!t]
	\centering
	\includegraphics[width=0.7\textwidth]{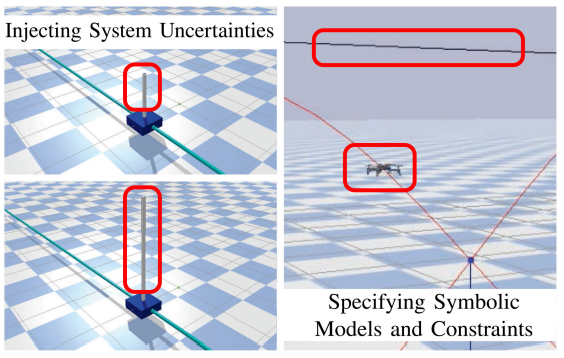}
	\caption{Safe-Control-Gym: adding constraints and injecting disturbances \cite{9849119}}
	\label{fig:Safe_Control_Gym}
\end{figure}

In Cyber-Physical Systems (CPS), Lyapunov and barrier functions contribute significantly to maintaining system safety, stability, and reliability. Lyapunov functions have been successfully applied to maintain stability under load disturbances and to mitigate risks due to cyber-attacks. Barrier functions protect against unsafe operating states, particularly during contingencies such as False Data Injection (FDI) attacks \cite{10252253}. Barrier functions integrated with RL methods ensure safe control in chemical processes by maintaining operational constraints, such as pressure, temperature, and flow rates. Lyapunov stability techniques ensure convergence of RL-based controllers to safe equilibrium \cite{yao2025lyapunov}.

In medical applications, guaranteeing safety is paramount. Lyapunov and barrier functions ensure compliance with critical constraints and improve patient outcomes. Safe RL leveraging Lyapunov functions ensures stability and safety in automated drug delivery, for instance, insulin infusion in diabetic patients. It guarantees that RL-based controllers maintain blood glucose levels within a safe operational range, effectively preventing adverse events \cite{yu2025deep}. Barrier functions have been proposed to ensure surgical robots avoid hazardous regions or unsafe interactions with critical tissues, making robotic surgery safer and more reliable \cite{fan2024learn}.

In intelligent transportation systems, Lyapunov stability criteria have been employed to optimize urban traffic signals to reduce congestion while guaranteeing minimal waiting times and avoiding unsafe traffic configurations \cite{wei2021recent}. Barrier functions ensure safety-critical constraints such as speed limits and collision avoidance in railway networks. These applications minimize operational risks while optimizing performance and efficiency \cite{lovetei2022environment}.

Aerospace systems rely heavily on guaranteed safety and stability provided by Lyapunov and barrier function methodologies. Safe RL augmented with CLF has been employed to optimize satellite trajectory planning, ensuring orbital stability and collision-free maneuvers \cite{yu2024safe, yi2025free}. Barrier functions in RL frameworks ensure adherence to critical safety constraints (e.g., angle-of-attack limits, airspeed criteria etc.) while maintaining optimal performance during aircraft operations \cite{jiang2023safely}.

Most literature in this review focuses on use of safe RL in the robotics domain, this is primarily because mobile robotic systems require satisfying hard constraints and computing a stabilizing policy for general applications which cannot be guaranteed with vanilla RL. Furthermore, safe RL methods provide an opportunity for online learning in case of using QP-based formulations, as discussed in~\cite{cai2021safe,emam2021safe,ma2021model}. A key domain that can benefit from safe RL with theoretical guarantees is power system control, which requires the RL algorithm to satisfy hard constraints for system states and dynamics~\cite{chen2022reinforcement}. There is limited literature considering use of Lyapunov and barrier functions for safe RL in power systems. The most notable approach~\cite{vu2021barrier} considers the use of barrier functions for emergency control during undervoltage load shedding. Overall, nonlinear dynamical systems which require satisfying hard constraints for operation can greatly benefit from using safe RL. 
%%%%%%%%%%%%%%%%%%%%%%%%%%%%%%%%%%%%%%%%%%%%%%%%%%%%%%%%%%%%%%%%%%%%%%%%%%%%%%%%%%%%%%%%%%%%%%%%%%%%%%%%%%%%%%%%%%%
\section{Conclusion}\label{section:Conclusions}
Safe RL using Lyapunov and barrier functions provides a principled route for adding stability and constraint-satisfaction structure to learning-based control. Under appropriate assumptions on dynamics, certificate validity, model uncertainty, and optimization feasibility, these methods can provide stronger training-time and deployment-time safety guarantees than unconstrained RL. This review summarizes and compares Lyapunov-based, barrier-based, and combined Lyapunov--barrier approaches, emphasizing how certificates enter rewards, critics, losses, QP filters, verification loops, and constrained policy optimization. The literature is rapidly moving from purely model-based certificate enforcement toward hybrid and data-driven approaches, including uncertainty-aware barriers, learned Lyapunov critics, offline safe RL, finite/fixed-time ADP, and CLF--CBF/CLBF formulations. The remaining gap between theory and deployment is centered on scalable certificate discovery, robust generalization under distribution shift, data efficiency, partial observability, QP feasibility, and real-world validation.
%%%%%%%%%%%%%%%%%%%%%%%%%%%%%%%%%%%%%%%%%%%%%%%%%%%%%%%%%%%%%%%%%%%%%%%%%%%%%%%%%%%%%%%%%%%%%%%%%%%%%%%%%%%%%%%%%%%

\section*{Declarations}

Please find below itemized declarations:
\begin{itemize}
	\item Funding: \em No, this research did not receive funding. \em
	\item Conflict of interest/Competing interests: \em No, I declare that the authors have no competing interests as defined by Springer, or other interests that might be perceived to influence the results and/or discussion reported in this paper.\em
	\item Ethics approval and consent to participate: \em I have read the Nature Portfolio journal policies on author responsibilities and submit this manuscript in accordance with those policies.\em
	\item Consent for publication: \em I agree to pay the APC in full if my article is accepted for publication (unless it is covered by an institutional agreement or journal partner, or a full waiver has been granted).\em
	\item Data availability: \em No / Not applicable. This manuscript does not report data generation or analysis.\em 
	\item Materials availability: \em No, all of the material is owned by the authors and/or no permissions are required.\em
	\item Code availability: \em No / Not applicable. This manuscript does not use code. \em
	\item Author contribution: \em D.S.K. and Z.A.B. wrote the main manuscript text and D.S.K. prepared all figures and tables (present at the end of the manuscript). All authors reviewed the manuscript.\em
\end{itemize}





\bibliography{bibliography}% common bib file

\end{document}